\documentclass[lettersize,journal]{IEEEtran}
\usepackage{amsmath,amsfonts}
\usepackage{algorithmic}
\usepackage{algorithm}
\usepackage{array}
\usepackage[caption=false,font=normalsize,labelfont=sf,textfont=sf]{subfig}
\usepackage{textcomp}
\usepackage{tcolorbox}
\usepackage{stfloats}
\usepackage{url}
\usepackage{bm}
\usepackage{amssymb}
\usepackage{verbatim}
\usepackage{graphicx}
\usepackage{cite}
\usepackage{stfloats}
\newtheorem{lemma}{\bf{\textit{Lemma}}}[section]
\newtheorem{theorem}{\bf{\textit{Theorem}}}[section]
\usepackage{hyperref}
\begin{document}

\title{Secure Analog Beamforming for Multi-user MISO Systems with Movable Antennas}

\author{Weijie Xiong, Jingran Lin, Kai Zhong, Liu Yang, Hongli Liu, Qiang Li, and Cunhua Pan
\thanks{This work was supported in part by the Natural Science Foundation of China (NSFC) under Grant 62171110. \textit{(Corresponding author: Jingran Lin, Cunhua Pan)}.}
\thanks{Jingran Lin and Qiang Li are with the School of Information and Communication Engineering, University of Electronic Science and Technology of China, Chengdu 611731, China, the Laboratory of Electromagnetic Space Cognition and Intelligent Control, Beijing 100083, China, and also with the Tianfu Jiangxi Laboratory, Chengdu, Sichuan 641419, China (e-mail: jingranlin@uestc.edu.cn; lq@uestc.edu.cn).}
\thanks{Weijie Xiong, Kai Zhong, Liu Yang, Hongli Liu are with the School of Information and Communication Engineering, University of Electronic Science and Technology of China, Chengdu 611731, China (e-mail: 202311012313@std.uestc.edu.cn; 201921011206@std.uestc.edu.cn; yangliu991022@gmail.com; hongliliu@std.uestc.edu.cn).}
\thanks{Cunhua Pan is with the National Mobile Communications Research Laboratory, Southeast University, China (e-mail: cpan@seu.edu.cn).}
}

\markboth{Journal of \LaTeX\ Class Files,~Vol.~14, No.~8, August~2021}%
{Shell \MakeLowercase{\textit{et al.}}: A Sample Article Using IEEEtran.cls for IEEE Journals}


\maketitle

\begin{abstract}
Movable antennas (MAs) represent a novel approach that enables flexible adjustments to antenna positions, effectively altering the channel environment and thereby enhancing the performance of wireless communication systems. However, conventional MA implementations often adopt fully digital beamforming (FDB), which requires a dedicated RF chain for each antenna. This requirement significantly increase hardware costs, making such systems impractical for multi-antenna deployments. To address this, hardware-efficient analog beamforming (AB) offers a cost-effective alternative. This paper investigates the physical layer security (PLS) in an MA-enabled multiple-input single-output (MISO) communication system with an emphasis on AB. In this scenario, an MA-enabled transmitter with AB broadcasts common confidential information to a group of legitimate receivers, while a number of eavesdroppers overhear the transmission and attempt to intercept the information. Our objective is to maximize the multicast secrecy rate (MSR) by jointly optimizing the phase shifts of the AB and the positions of the MAs, subject to constraints on the movement area of the MAs and the constant modulus (CM) property of the analog phase shifters. This MSR maximization problem is highly challenging, as we have formally proven it to be NP-hard. To solve it efficiently, we propose a penalty constrained product manifold (PCPM) framework. Specifically, we first reformulate the position constraints as a penalty function, enabling unconstrained optimization on a product manifold space (PMS), and then propose a parallel conjugate gradient descent algorithm to efficiently update the variables. Simulation results demonstrate that MA-enabled systems with AB can achieve a well-balanced performance in terms of MSR and hardware costs.
\end{abstract}

\begin{IEEEkeywords}
Movable antennas, physical layer security, analog beamforming, constant modulus, penalty constrained product manifold.
\end{IEEEkeywords}

\section{Introduction}
\IEEEPARstart{W}{ireless} communication networks are inherently vulnerable to information leakage due to their open-access nature, making them susceptible to eavesdropping attacks \cite{khisti2010secure}. To address this challenge, the concept of physical layer security (PLS) was introduced, which leverages spatial diversity to enhance communication for legitimate users (LUs) while obstructing eavesdroppers (EVEs) \cite{6772207}. Over the years, significant progress has been made in PLS, driven by innovative techniques such as reconfigurable intelligent surfaces (RIS) \cite{11095722,10922160,10922191}, frequency diverse array (FDA)-based beamforming \cite{lin2017physical,akkoc2023time,nusenu2020authentication}, and artificial noise (AN)-aided designs \cite{al2018beamforming,zhang2018artificial,chu2022joint}, which generally exploit the statistical independence between LU and EVE channels to optimize LU reception while degrading EVE interception performance.

Despite advancements in PLS, sophisticated interception strategies, such as deploying multiple EVEs \cite{8873672,lin2018robust} and leveraging associations with low-priority LU \cite{lin2017physical,zhang2019transmit}, pose increasing challenges to secure communication. The presence of multiple EVEs exacerbates the correlation between legitimate and eavesdropping channels, particularly in densely deployed or multi-user environments \cite{li2011multicast}, which undermines the spatial diversity essential for PLS. Traditional techniques constrained by fixed-position antennas (FPAs) offer limited solutions to address this issue, even as the number of antennas increases \cite{ragheb2023ris}. Consequently, these approaches often struggle to sufficiently differentiate LU and EVE channels, resulting in degraded secrecy performance.

To address these challenges, the novel technology of movable antennas (MAs) offers a promising solution. Unlike conventional FPA-based communication systems, MAs overcome performance limitations by leveraging the mobility of antennas to enhance system adaptability \cite{zhu2023movable, zheng2024flexible}. In MA-enabled systems, antennas are connected to radio frequency (RF) chains via flexible cables, allowing dynamic repositioning \cite{chen2023joint}. This mobility enables MAs to fully exploit spatial degrees of freedom (DoFs), facilitating channel reconstruction and significantly improving communication performance. Specifically, MAs can reposition themselves to decouple LU and EVE channels, thereby significantly enhancing secrecy performance. Applications, hardware architectures, and channel characterizations of MA-enabled communication systems are extensively discussed in \cite{zhu2023movable, zheng2024flexible}, while joint designs for transmit covariance and antenna movement in MIMO systems are detailed in \cite{ma2023mimo, chen2023joint, zhu2023modeling}. 

Building upon the benefits of MAs for wireless communication systems \cite{zhu2023movable, zheng2024flexible, ma2023mimo, chen2023joint, zhu2023modeling}, MAs have also been applied to PLS to enhance performance \cite{hu2024secure,cheng2024enabling,tang2024secure,ding2024movable,ma2025movable,cheng2025movable}. For example, \cite{hu2024secure} first considered a transmitter with multiple movable antennas constrained to one-dimensional motion communicating with a single-antenna LU in the presence of an EVE. The secrecy rate is maximized by jointly optimizing the transmit beamforming and antenna positions under transmit-power and movement constraints. Building on this, \cite{cheng2024enabling} further investigated MA-enabled PLS systems, where MAs are capable of two-dimensional movement, offering additional DoFs to improve system performance. In \cite{tang2024secure}, the authors extended the study to MIMO systems by considering a multi-antenna EVE and a multi-antenna LU. Moreover, \cite{ding2024movable} extended the analysis to multi-user cases by considering multiple BOBs to mitigate the cooperative interception of multiple EVEs. Furthermore, MAs have been integrated into emerging systems to improve PLS, such as integrated sensing and communication (ISAC) \cite{ma2025movable} and frequency diverse array (FDA) systems \cite{cheng2025movable}. Although these studies demonstrated substantial performance improvements, they predominantly focused on scenarios involving either a single LU \cite{hu2024secure,cheng2024enabling,tang2024secure} or multiple LUs \cite{ding2024movable,ma2025movable,cheng2025movable} receiving distinct messages, i.e., unicasting. In contrast, multicasting scenarios \cite{li2023joint,sun2022secure}, where a common confidential message is simultaneously delivered to multiple LUs, represent an important yet relatively unexplored area for MA-enabled PLS.

Moreover, prior studies consistently indicate that increasing the number of antennas in MA-enabled systems provides additional DoFs, thereby significantly reducing channel correlation between LUs and EVEs, thus enhancing communication security \cite{hu2024secure, cheng2024enabling, tang2024secure, ding2024movable, ma2025movable, cheng2025movable}. Nonetheless, practical implementation of systems with numerous antennas remains challenging due to increased hardware costs. These challenges primarily arise because MA systems typically employ fully digital beamforming (FDB) architectures, which require a dedicated RF chain for each antenna to achieve precise beamforming. Therefore, there is a pressing need to develop cost-effective solutions that can sustain high performance.

Recently, analog beamforming (AB), which can be implemented using low-cost phase shifters and a single variable gain amplifier (VGA), has emerged as a promising solution for cost-effective deployment of antenna arrays. Unlike FDB, which requires a dedicated RF chain for each antenna element, AB leverages analog phase shifters (PSs) and a single VGA to control the beamforming vector. This approach constrains the beamforming vector to constant modulus (CM) entries, with the magnitude of each entry determined by the gain provided by the VGA. Consequently, AB achieves a low peak-to-average ratio (PAR) for transmitted signals, enabling the use of hardware-efficient nonlinear PSs at the transmitter. Due to its potential to reduce hardware complexity, AB has been extensively studied for improving wireless communication performance. Previous studies have addressed various aspects of AB design, including interference suppression, spectral efficiency, and energy efficiency \cite{mohammed2012single, pan2014constant, zhang2017constant}; however, these works did not consider PLS. In the context of PLS, \cite{xiong2024constant} proposed a two-stage optimization framework to design AB for both information-bearing signals and AN. Additionally, \cite{zhao2016phase} investigated a phase-only zero-forcing (ZF) precoding approach for large-scale antenna arrays. More recently, \cite{zhu2016secure} characterized an ergodic secrecy rate lower bound under AB and hybrid beamforming architectures.

MAs have demonstrated significant effectiveness in enhancing PLS, particularly in scenarios involving multiple LUs. However, as the number of antennas increases to manage more complex communication environments, the associated hardware costs present considerable challenges. These challenges naturally encourage the adoption of AB, a hardware-efficient solution. This paper investigates the potential of AB to further enhance PLS in MA-enabled systems, highlighting its architectural and design advantages in reducing costs while preserving secure communication performance. The key contributions are as follows:
\begin{itemize} 
\item We investigate PLS in MA-enabled multiple-input single-output (MISO) systems with AB. The goal is to maximize the multicast secrecy rate (MSR) by jointly optimizing AB phase shifts and MA positions under movement area and CM constraints. While existing works \cite{cheng2024enabling, hu2024secure} have addressed PLS in MA systems, they primarily focus on unicasting scenarios, overlooking multicasting environments. Moreover, prior designs rely on FDB, where secure beamforming can be solved via generalized eigendecomposition under total power constraints. In contrast, CM constraints render the principal eigenvector suboptimal or infeasible, and naive projection onto the CM set leads to significant secrecy performance loss, making the problem highly challenging.
\item We rigorously prove that the formulated problem is NP-hard, highlighting its computational complexity and the challenges associated with finding optimal solutions. In light of the above difficulties, we focus on developing some tractable approaches to finding high-quality approximate solutions to the problem. 
\item  We observe that the complex circle manifold (CCM) naturally satisfies the CM constraints in AB. To strictly enforce the inequality constraints on MA positions, we adopt a penalty function method to handle violations. Building on this foundation, we propose a penalty constrained product manifold (PCPM) framework. By transforming the inequality constraints into a penalty function using smoothing techniques, the problem is reformulated as an unconstrained optimization on the product manifold space (PMS). A parallel conjugate gradient descent algorithm is then developed to efficiently update variables on the PMS. We prove that the proposed PCPM framework converges to a KKT point. 
\end{itemize}

The remainder of this paper is organized as follows: Section II introduces the system model and problem statement. In Section III, we reformulate the problem and propose a PCPM framework to solve it. Simulation results are presented in Section IV, and Section V concludes the paper.

The following notations are used throughout the paper. A vector and a matrix are represented by $\bf a$ and $\bf A$ respectively; $(\cdot)^T$, $(\cdot)^H$ and $(\cdot)^*$ denote the transpose, conjugate transpose and conjugate respectively. $\bf I$ denotes an identity matrix with an appropriate dimension; ${\mathbb C}^N$ denotes the set of complex vectors of dimension $N$; the circularly symmetrix complex Gaussian distribution with mean $\mu $ and variance $\sigma^2 $ is denoted as $\mathcal{C N}(\mu, \sigma^2)$; $\text{Tr}({\bf A})$,  $||\cdot||_F$, $||\cdot||_2$ and  $|\cdot|$ represents trace operator, Frobenius norm, Euclidean norm and absolute value; $\text{diag}({\bf a})$ represnts a diagonal matrix where the elements of the vector $\bf a$ are placed on the main diagonal; the phase of each element of a matrix is denoted as $\text{arg} ({\bf A})$; $\Re\{\cdot\}$ denotes the real part of a complex matrix; ${\bf A}\odot{\bf B}$ represents Kronecker product.

\section{System Model and Problem Formulation}
\label{sec:format}

\begin{figure}[!htbp]
  \begin{center}
  \includegraphics[width=2.5in]{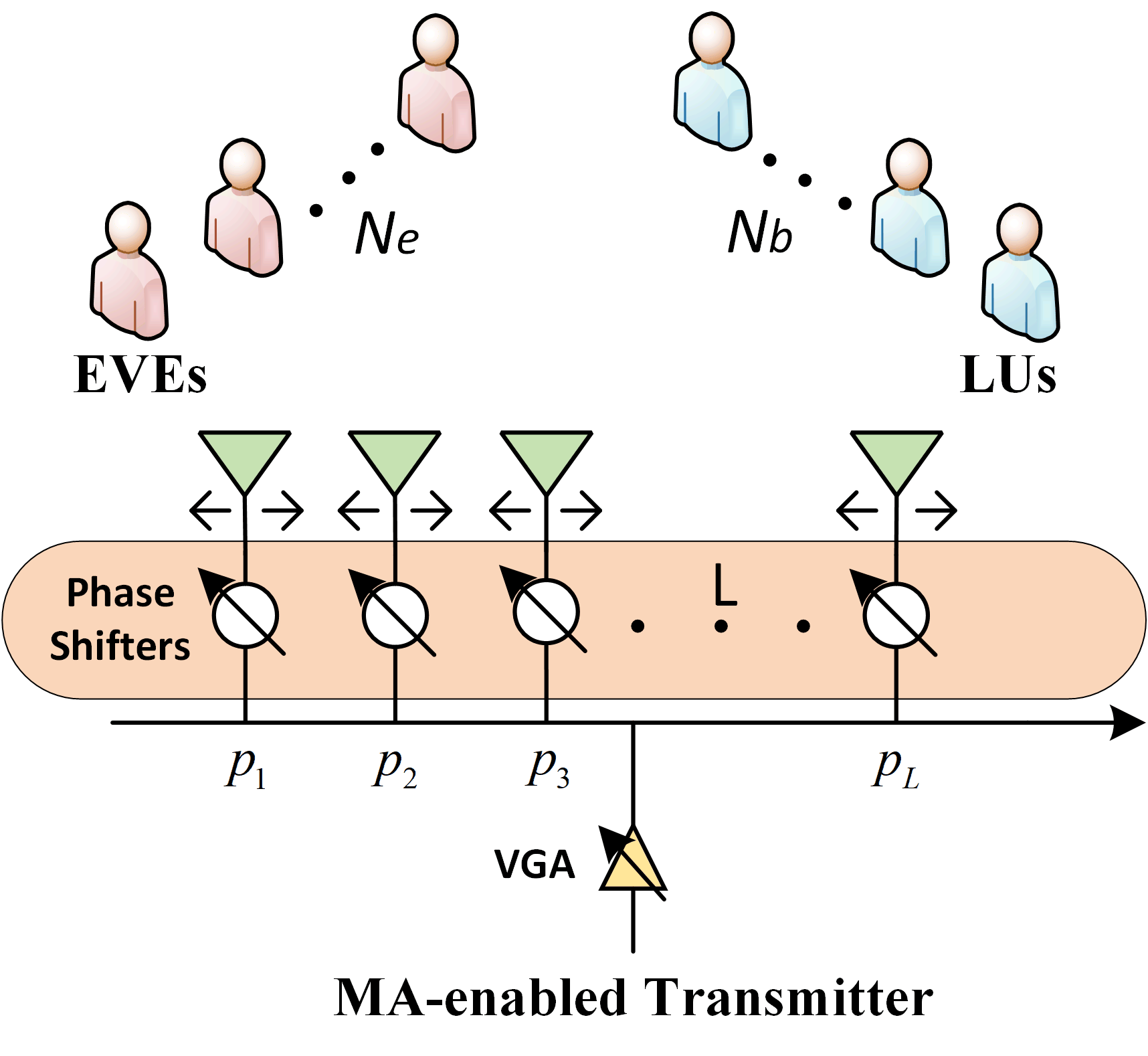}\\
  \caption{The MA-enabled MISO communication system with AB.}\label{systemmodel}
  \end{center}
\end{figure}

As illustrated in Fig. \ref{systemmodel}, we consider a multiuser MISO downlink network, where the transmitter employs analog PSs driven by a common VGA to broadcast confidential information to \( N_b \) LUs, in the presence of \( N_e \) EVEs attempting to intercept the confidential data. Each of the \( N_b \) LUs and \( N_e \) EVEs is equipped with a single, fixed-position antenna, while the transmitter is equipped with a linear MA array of size \( L \). The position of the \( l \)-th antenna at the transmitter is denoted by \( p_l, \ 1 \le l \le L \), and the positions of the \( L \) antennas can be compactly expressed as \( {\bf p} = [p_1, p_2, \dots, p_L]^T \in \mathbb{R}^L \).

Let $ s \in \mathbb{C}$ be the coded conﬁdential information for LUs with unit power. The transmit signal is given by,
\begin{equation}
{\bf x}=\sqrt {{P}}{\bf w}s {\in \mathbb{C}^{L}},
\end{equation}
where $ \sqrt {{P}} > 0$ represents the transmit power and $\mathbf{w} \in \mathbb{C}^{L}$ is the transmit AB controlled by the PSs with CM, i.e.,
\begin{equation}
|w_l| = 1,\quad \forall l = 1,...,L, \label{beamformingw}
\end{equation}
Then, the received signal at the $b$-th LU and the $e$-th EVE are respectively given by,
\begin{subequations}
\begin{align}
& {\hat y}_{b}({\bf w},{\bf p})=\sqrt {{P}}\mathbf{h}_b^H({\bf p}) \mathbf{w}s +{\hat{n}}_{b} \in \mathbb{C},\quad \forall b=1,...,N_b,\\
& {\check y}_e({\bf w},{\bf p})= \sqrt {{P}}{\bf{g}}_e^H({\bf p}) {\bf{w}}s +{\check{n}}_e \in \mathbb{C},\quad \forall e=1,...,N_e.
\end{align}
\end{subequations}
where ${\hat{n}}_{b} \sim \mathcal{C N}({0}, {\hat\sigma}_b^2)$ and ${\check{n}}_e \sim \mathcal{C N}({0}, {\check\sigma}_e^2)$ represent the Gaussian noise for the $b$-th LU and the $e$-th EVE, respectively. $\mathbf{h}_b({\bf p})$ and $\mathbf{g}_e({\bf p})$ denote the legitimate channels from the Alice to the $b$-th LU and $e$-th EVE, respectively.

In an MA-enabled network, the channel vector depends not only on the propagation environment but also on the positions of the MAs. In this paper, we adopt a field-response-based channel model \cite{zhu2023movable}, which assumes that the size of the transmit array is much smaller than the signal propagation distance, satisfying the far-field assumption \cite{zheng2024flexible}. Under this assumption, as the MAs move, the signal for each propagation path undergoes only phase variations, while the angle of departure (AoD) and amplitude remain unchanged \cite{ma2023mimo, chen2023joint}. We further assume that the channel from the MAs to each LU and EVE experiences the same total number of paths, denoted as $M_p$. Based on these assumptions, the spatial AoDs for the $m$-th propagation path to the $b$-th LU and the $e$-th EVE are denoted as ${\hat \theta_{b,m}} \in \left[-\frac{\pi}{2}, \frac{\pi}{2}\right]$ and ${\check \theta_{e,m}} \in \left[-\frac{\pi}{2}, \frac{\pi}{2}\right]$, respectively, where $1 \le m \le M_p$. Then, the field response vectors (FRVs) for the $m$-th propagation path to the $b$-th LU and $e$-th EVE are given by,
\begin{subequations}
\begin{align}
& {\bf a}({\bf p},\hat\theta_{b,m}) =  [e^{j\tfrac{2\pi }{\lambda }{ p_1}{\cos}\hat \theta_{b,m}  },...,e^{j\tfrac{2\pi}{\lambda }{ p_L}{\cos}\hat\theta_{b,m}  }]^T\in\mathbb{C}^{L},\\
& {\bf a}({\bf p},\check\theta_{e,m}) =  [e^{j\tfrac{2\pi }{\lambda } { p_1}{\cos}\check\theta_{e,m} },...,e^{j\tfrac{2\pi}{\lambda }{ p_L}{\cos}\check\theta_{e,m}  }]^T\in\mathbb{C}^{L},
\end{align}
\end{subequations}
where $\lambda$ is the wavelength. Let $\hat{\beta}_{b,m}$ and $\check{\beta}_{e,m}$ represent the complex path gains for the $m$-th path to the $b$-th LU and $e$-th EVE, respectively. Hence, the communication channel between the MAs and the $b$-th LU and $e$-th EVE, considering a total of $M_p$ paths, can be expressed as,
\begin{subequations}
\begin{align}
  &\mathbf{h}_b({\bf p})=\sum_{m=1}^{M_p}{\hat\beta_{b,m}}{\bf a}({\bf p},\hat\theta_{b,m})  \in \mathbb{C}^L,\quad \forall b=1,...,N_b,\\ &\mathbf{g}_e({\bf p})=\sum_{m=1}^{M_p}{\check\beta_{e,m}}{\bf a}({\bf p},\check\theta_{e,m})\in \mathbb{C}^L,\quad \forall e=1,...,N_e.  
\end{align}
\label{ffrmodel}%
\end{subequations}

\textit{\textbf{Remark 1:}} As demonstrated in \cite{zhu2023modeling}, the field-response channel model in (\ref{ffrmodel}) can naturally degenerate into spatially-correlated channel models, such as Rayleigh and Rician fading, under the condition of an infinite number of propagation paths with random path gains. In scenarios characterized by rich scattering, which are typical in dense networks, spatially-correlated channel models can therefore be considered as special cases of the field-response model \cite{zhu2025tutorial}. This implies that the field-response model inherently captures channel correlations that may influence secrecy performance in such environments.

\textit{\textbf{Remark 2:}} In this paper, we assume that the transmitter has perfect channel state information (CSI) of EVEs, which is possible, say, EVEs are also users of the system, and the transmitter aims to provide different types of users with different services. The services should be provided exclusively for the target users. Moreover, for active EVEs, the CSI can be estimated from the EVEs transmission. More interestingly, even for a passive EVE, there is a possibility for one to estimate the CSI through the local oscillator power inadvertently leaked from the EVEs receiver RF frontend \cite{mukherjee2012detecting}. Besides, the impact of imperfect CSI on the considered system will be evaluated via simulations in Section IV.

According to the information-theoretic principles of secure communication \cite{liang2009information}, the PLS can be characterized by the MSR, defined as the positive difference between the achievable rates of LUs and EVEs, given by,
\begin{equation}
{R_{sec}({\bf w},{\bf p})}= \mathop {\min }_{ b=1,...,N_b \atop e=1,...,N_e}  \left[   {\hat R_b({\bf w},{\bf p})}- {\check R_e({\bf w},{\bf p})}   \right]^{+},\label{MCSR}
\end{equation}
where $[u]^{+} \triangleq \max (u, 0)$ indicates the maximum number between 0 and $u$, and,
\begin{subequations}
\begin{align}
 {\hat R_b({\bf w},{\bf p})} &= \log(1+  |\sqrt {{P}}{\bf h}_b^H({\bf p}) \mathbf{w}|^2/{{\hat\sigma}_b^2}),  \\
 {\check R_e({\bf w},{\bf p})}& = \log(1+  {|\sqrt {{P}}{\bf g}_e^H({\bf p}) \mathbf{w}|^2}/{{\check\sigma}_e^2}), 
\end{align}
\end{subequations}
are the achievable rate for $b$-th LU and $e$-th EVE, respectively.

\textit{\textbf{Remark 3:}} Our work addresses a multicast scenario where a common confidential message is sent to all LUs. This model is a direct consequence of the considered hardware-efficient AB architecture (see Fig. \ref{systemmodel}). In this setup, a single VGA and a network of PSs are driven by a single data stream, making multicasting a natural application for this low-cost design. Supporting distinct messages for different users would require a more complex hybrid beamforming architecture with multiple RF chains \cite{10922191}, which entails significantly higher hardware costs and complexity.

Our goal is to maximize ${R_{sec}({\bf w},{\bf p})}$, by jointly optimizing the AB $\bf w$ at the transmitter and the positions of the MAs $\bf p$. Hence, the optimization problem is formulated as,
\begin{subequations}
\begin{align}
 \max _{{\bf w}, {\bf p}} \quad & {R_{sec}({\bf w},{\bf p})}, \label{Rssecret} \\
 \text { s.t. } \quad & \left|w_l\right|=1, \quad l=1, \ldots, L, \label{CMC} \\
&  p_{l+1} - p_{l} \ge \frac{\lambda}{2}, \quad 1 \le l \le L-1, \label{minidistance}\\
&  p_{1} \ge 0 ,\quad p_{L}  \le D. \label{antennaselect}
\end{align}
\label{objectivefunction}%
\end{subequations}
where constraints (\ref{minidistance}) ensure that the distance between any two MAs is no smaller than $\tfrac{\lambda}{2}$ to avoid the coupling effect; constraint (\ref{antennaselect}) guarantees that the position of any MA is no greater than $D$ and no smaller than 0. Note that to make (\ref{minidistance}) always hold, it is clear to determine that $D$ in (\ref{antennaselect}) should be no smaller than $\frac{\lambda}{2}(L - 1)$. 

Due to the non-smoothness and non-concavity of the objective function (\ref{Rssecret}), the non-convex CM constraints (\ref{CMC}), and the non-convex coupled-variable constraints (\ref{minidistance}), problem (\ref{objectivefunction}) becomes a highly non-convex problem, as highlighted by the following fact.

\begin{lemma} \label{lemma1} \itshape
Problem (\ref{objectivefunction}) is NP-hard in general.
\end{lemma}

\quad {\bf\textit{Proof}}{\bf:} See Appendix \ref{appendixA}. $\hfill\blacksquare$

In light of the above difﬁculties, in the following sections we focus on developing some tractable approaches to ﬁnding high-quality approximate solutions to problem (\ref{objectivefunction}). We will ﬁrst reformulate problem (\ref{objectivefunction}) into an easier-to-handle form, and then the penalty constrained product manifold (PCPM) framework is proposed to solve the reformulated problem efficiently.

\textit{\textbf{Remark 4:}} Although both MAs and AB involve phase manipulation, they operate on different aspects of the system and are therefore complementary. MAs provide geometric channel shaping: by physically repositioning antenna elements, they modify the array steering vectors, reshape ${\mathbf{h}_b(\mathbf{p}),\mathbf{g}_e(\mathbf{p})}$, and reduce LU–EVE channel correlation, subject to spatial constraints on antenna placement. AB provides electrical beam shaping: with constant modulus phase shifters, it cannot alter the underlying channels (and thus cannot directly reduce correlation), but it efficiently enhances LUs’ achievable rates and forms deep nulls toward EVEs on the given channels. Moreover, it achieves this using a single RF chain with one VGA, thereby avoiding the RF-chain overhead of FDB systems. Joint optimization of MA positions and AB phases allows MAs to create a more separable channel landscape that AB then exploits, resulting in higher MSR and improved robustness compared with either technique alone or MA designs relying solely on FDB.

\section{Secrecy rate Maximization Scheme for MA-enabled system with AB}
In this section, we propose a framework for MSR maximization. First, we analyze and reformulate the NP-hard problem, transforming it into a smooth problem with inequality constraints. Then, we introduce the PCPM framework to solve it.

\subsection{Problem Reformulation}
To reduce the difficulty of directly handling the objective function (\ref{Rssecret}), we first manage it by dropping the log operation and then it can be equivalently written as,
\begin{equation}
\begin{aligned}
\max _{{\bf w}, {\bf p}} \quad & \max \left\{ \frac{ \min\limits_{b=1,...,N_b} 1+P|{\bf h}_b^H({\bf p}) \mathbf{w}|^2/{\hat\sigma}_b^2 }{\max\limits_{e=1,...,N_e} 1+P|{\bf g}_e^H({\bf p}) \mathbf{w}|^2/{\check\sigma}_e^2} ,1 \right\}\\
\text{s.t.} \quad& \text{(\ref{CMC}), (\ref{minidistance}), and (\ref{antennaselect}).}
\end{aligned}
\end{equation}
Notice that the objective function \( R_{sec}({\bf w}, {\bf p}) \) is defined as the maximum value between the constant 1 and $\frac{ \min\limits_{b=1,...,N_b} 1+P|{\bf h}_b^H({\bf p}) \mathbf{w}|^2/{\hat\sigma}_b^2 }{\max\limits_{e=1,...,N_e} 1+P|{\bf g}_e^H({\bf p}) \mathbf{w}|^2/{\check\sigma}_e^2}$. Since only the fractional term depends on \( {\bf w} \) and \( {\bf p} \), the problem can initially focus on maximizing these terms. If its solution is less than 1, the secrecy rate is negative, making secure transmission unfeasible. This approach streamlines evaluating communication feasibility before tackling the full optimization problem. Thus, we consider the following maximization problem,
\begin{equation}
\begin{aligned}
\max _{{\bf w}, {\bf p}} \quad &  \frac{ \min\limits_{b=1,...,N_b} 1+P|{\bf h}^H_b({\bf p}) \mathbf{w}|^2/{\hat\sigma}_b^2 }{\max\limits_{e=1,...,N_e} 1+P|{\bf g}_e^H({\bf p}) \mathbf{w}|^2/{\check\sigma}_e^2} \\
\text{s.t.} \quad& \text{(\ref{CMC}), (\ref{minidistance}), and (\ref{antennaselect}).}
\end{aligned}
\label{reformualteobnew2}
\end{equation}
which can be re-expressed as the following form after changing the maximization to a minimization and also exchanging the numerator and the denominator in (\ref{reformualteobnew2}),
\begin{equation}
\begin{aligned}
\min _{{\bf w}, {\bf p}} \quad &  \frac{ \max\limits_{e=1,...,N_e} 1+P|{\bf g}_e^H({\bf p}) \mathbf{w}|^2/{\check\sigma}_e^2  }{\min\limits_{b=1,...,N_b} 1+P|{\bf h}_b^H({\bf p}) \mathbf{w}|^2/{\hat\sigma}_b^2 } \\
\text{s.t.} \quad& \text{(\ref{CMC}), (\ref{minidistance}), and (\ref{antennaselect}).}
\end{aligned}
\label{reformualteobnew3}
\end{equation}
However, although the above operation has simplified the difficulty of handling the objective function (\ref{Rssecret}) by disregarding the $[\cdot]^+$ and log operations, the reformulated objective function (\ref{reformualteobnew3}) is still non-smooth and non-convex. To turn problem (\ref{reformualteobnew3}) into a more manageable form, we smooth the objective of (\ref{reformualteobnew3}) by applying the following log-sum-exponential approximation \cite{boyd2004convex},

\begin{lemma} \label{lemma2} \itshape
Given $c_1,c_2,..., c_K \in \mathbb{R}$, it holds for any $\alpha > 0$ that,
\begin{subequations}
    \begin{align}
        \max_{k=1,...,K} c_k \le \alpha \log \sum_{k=1}^K \exp({\frac{c_k}{\alpha}}) \le \max_{k=1,...,K} c_k + \alpha \log K.
    \end{align}
\end{subequations}
Moreover, the inequalities become tight as $\alpha  \to  0$.
\end{lemma}

\quad {\bf\textit{Proof}}{\bf:} The proof can be found in \cite{nesterov2013introductory} and the detailed derivation is omitted for brevity. $\hfill\blacksquare$

Using Lemma \ref{lemma2}, problem (\ref{reformualteobnew3}) is smoothly approximated as,
\begin{equation}
\begin{aligned}
\min _{{\bf w}, {\bf p}} \quad &  \frac{ \alpha \log \sum\limits_{e=1}^{N_e} \text{exp}\left( \frac{1+{\check t_e}|{\bf g}_e^H({\bf p},\check\theta_e) \mathbf{w}|^2} {\alpha} \right)}{-\alpha \log \sum\limits_{b=1}^{N_b} \text{exp}\left( -\frac{ 1+{\hat t_b}|{\bf h}_b^H({\bf p},\hat\theta_b)\mathbf{w}|^2 }{\alpha} \right) } \\
\text{s.t.} \quad& \text{(\ref{CMC}), (\ref{minidistance}), and (\ref{antennaselect}).}
\end{aligned}
\label{reforset}
\end{equation}
where ${\check t_e}={P}/{{\check\sigma}_e^2}$ and ${\hat t_b}={P}/{{\hat\sigma}_b^2}$ are a constant for symbol simplification. Note that the reformulated problem (\ref{reforset}) is still challenging to solve, due to the non-convex nature of the CM constraints (\ref{CMC}) and the coupled-variable inequality constraints (\ref{minidistance}). We observe that the product manifold space (PMS) naturally satisﬁes the CM constraints (\ref{CMC}), and the inequality constraints (\ref{minidistance}) can be transformed into a non-negative smooth function. Leveraging these structural characteristics, we propose the PCPM framework to solve the reformulated problem (\ref{reforset}) as follows.  

\subsection{The proposed PCPM framework}
In this subsection, we drive the steps for solving problem (\ref{reforset}) by utilizing the PCPM framework. Specifically, we first convert the inequality constraints (\ref{minidistance}) and (\ref{antennaselect}) into a penalty function using smoothing techniques, enabling the reformulation of the problem as an unconstrained one on the PMS. To solve the reformulated unconstrained problem over the product manifold, an efficient conjugate gradient descent (CGD) algorithm is derived without relaxing the objective function.

\subsubsection{Penalty Inequality Constraints over Objective Function}
Generally, directly adding inequality constraints to the objective function may struggle to satisfy the constraints, leading to violations and potentially divergent behavior. A more accurate way for managing inequality constraints is the exterior penalty (EP) approach \cite{ruszczynski2011nonlinear}. Instead of incorporating constraints directly, this technique augments the objective function with an exterior penalty for constraint violations. Based on this, (\ref{reforset}) can be transformed into an exterior penalty function optimization problem with CM constraints given as,
\begin{subequations}
\begin{align}
\min _{{\bf w}, {\bf p}}  & \left\{\begin{array}{l} \frac{ \alpha \log \sum\limits_{e=1}^{N_e} \text{exp}\left( (1+{\check t_e}|{\bf g}_e^H({\bf p}) \mathbf{w}|^2) /\alpha \right)}{-\alpha \log \sum\limits_{b=1}^{N_b} \text{exp}\left( -( 1+{\hat t_b}|{\bf h}_b^H({\bf p})\mathbf{w}|^2 )/\alpha \right) }  \\+\rho  \textstyle\sum\limits_{l=1}^{L-1}   
 \max(0,{g_l(p_{l+1},p_{l})}) \\+ \rho\max(0,{f_1(p_{1})})+ \rho\max(0,{f_2(p_{L})}),  \end{array}\right\} \\
 \text { s.t. }  & \left|w_l\right|=1, \quad l=1, \ldots, L, 
\end{align}
\label{reforfirsm1}%
\end{subequations}
where $\rho > 0$ is the penalty factor, and,
\begin{subequations}
\begin{align}
& g_l(p_{l+1},p_{l}) = p_l - p_{l+1} + \frac{\lambda}{2},\quad \forall l \in [1, L-1], \label{exterg} \\
& f_1(p_{1}) = -p_1 ,\label{exterf1}\\
& f_2(p_{L}) = p_L - D. \label{exterf2}
\end{align}
\label{totalett}%
\end{subequations}
where (\ref{exterg}), (\ref{exterf1}) and (\ref{exterf2}) are EP functions for constraints (\ref{minidistance}) and (\ref{antennaselect}), respectively. Note these EP functions are also non-smooth and can be challenging to solve directly. Fortunately, they all comprise a two-term maximum, which can similarly be smoothed by applying the following log-sum-exponential approximation according to Lemma \ref{lemma2}. Based on this fact, (\ref{reforfirsm1}) can be transformed as a smooth EP function optimization problem with CM constraints,
\begin{subequations}
\begin{align}
\min _{{\bf w}, {\bf p}}  & \left\{\begin{array}{l} \frac{ \alpha \log \sum\limits_{e=1}^{N_e} \text{exp}\left( (1+{\check t_e}|{\bf g}_e^H({\bf p}) \mathbf{w}|^2) /\alpha \right)}{-\alpha \log \sum\limits_{b=1}^{N_b} \text{exp}\left( -( 1+{\hat t_b}|{\bf h}_b^H({\bf p})\mathbf{w}|^2 )/\alpha \right) }  \\+\rho\gamma   \textstyle\sum\limits_{l=1}^{L-1}  
 \log(1+e^{g_l(p_{l+1},p_{l})/\gamma }) \\+ \rho\gamma\log(1+e^{f_1(p_{1})/\gamma})+ \rho\gamma\log(1+e^{f_2(p_{L})/\gamma}),  \end{array}\right\} \label{reforfirsm2obj} \\
 \text { s.t. }  & \left|w_l\right|=1, \quad l=1, \ldots, L, \label{consnn}
\end{align}
\label{reforfirsm2}%
\end{subequations}
Note that the problem (\ref{reforfirsm2}) involves CM constraints that are challenging to express and handle in a linear space. Manifolds, being ﬂexible and capable of capturing non-linear relationships, offer a natural framework to express and incorporate these constraints. The manifold’s local Euclidean nature allows it to express complex relationships in a way that aligns with the underlying geometry.

\subsubsection{Construction of the PMS}
In this subsection, we construct a PMS to satisfy the constraints of $\bf w$ and ${\bf p}$. Ensuring adherence to these constraints is vital for the effectiveness and reliability of the PMS.

Constraint (\ref{consnn}) satisﬁes the complex circle manifold ${\cal M}_{\bf w}$ \cite{10801197},
\begin{equation}
\mathcal{M}_{\bf w}=\left\{{\bf w} \in \mathbb{C}^{L} | \,|w_l| =1, \quad l=1, \ldots, L \right\}. \label{manifoldw}
\end{equation}

Variable ${\bf p}\in\mathbb{R}^L$ satisﬁes the Euclidean space manifold ${\cal M}_{\bf p}$,
\begin{equation}
\mathcal{M}_{\bf p}=\left\{{\bf p} \in \mathbb{R}^{L} \right\}. \label{manifoldf}
\end{equation}

To represent the feasible set of solutions, we construct the PMS as the Cartesian product of those individual manifolds in (\ref{manifoldw}) and (\ref{manifoldf}). The PMS is given by,
\begin{equation}
\mathcal{M}=\mathcal{M}_{\bf w} \times \mathcal{M}_{\bf p}=\left\{({\bf w}\in \mathbb{C}^{L}, {\bf p}\in \mathbb{R}^{L}): |w_l| =1, \forall l, {\bf p}  \right\}.   \label{productmanifold} 
\end{equation}

Based on (\ref{productmanifold}), Problem (\ref{reforfirsm2}) can be reformulated as an unconstrained problem over PMS, denoted as,
\begin{equation}
\min _{({\bf w}, {\bf p})\in\mathcal{M}} \phi({\bf w}, {\bf p})=\left\{\begin{array}{l} \frac{ \alpha \log \sum\limits_{e=1}^{N_e} \text{exp}\left( {(1+{\check t_e}|{\bf g}_e^H({\bf p}) \mathbf{w}|^2)}/{\alpha} \right)}{-\alpha \log \sum\limits_{b=1}^{N_b} \text{exp}\left( {-( 1+{\hat t_b}|{\bf h}_b^H({\bf p})\mathbf{w}|^2 )}/{\alpha} \right) }  \\+\rho\gamma   \textstyle\sum\limits_{l=1}^{L-1}  
 \log(1+e^{g_l(p_{l+1},p_{l})/\gamma }) \\+ \rho\gamma\log(1+e^{f_1(p_{1})/\gamma})\\+ \rho\gamma\log(1+e^{f_2(p_{L})/\gamma}),  \end{array}\right\}
\label{reforoverpro}    
\end{equation}

The manifold space of (\ref{reforoverpro}) introduces curvature and nonlinearity, presenting challenges for algorithm design in PMS directly. To overcome this challenge, it is essential to construct the uniﬁed tangent space (UTS), which acts as a local linear approximation to the manifold. This facilitates the extension of traditional optimization algorithms, originally designed for ﬂat spaces, to navigate and optimize over the curved surfaces of the manifold.

In the case of the complex circle manifold  $\mathcal{M}_{\bf w}$ in (\ref{manifoldw}), the tangent space at any point ${\bf w} \in \mathbb{C}^L $ is the complex hyperplane orthogonal to ${\bf w}$, given by,
\begin{equation}
    \mathrm{T}_{\mathbf{w}} \mathcal{M}_{\bf w}=\left\{{\bm \xi}_{\mathbf{w}} \in \mathbb{C}^{L}: \Re\left\{{\bm \xi}_{\mathbf{w}} \odot \mathbf{w}^*\right\}=\mathbf{0}_{L}\right\}, \label{tsw}
\end{equation}
where ${\bm \xi}_{\bf w}$ is the tangent vector at point ${\bf w}$. This condition ensures that the tangent vector remains orthogonal to ${\bf w}$ thus constraining the vector to lie within the manifold and avoiding any radial components. Geometrically, this tangent space forms a complex hyperplane, representing all possible directions of movement that stay tangential to the complex circle manifold at ${\bf w}$.

In the case of the Euclidean space manifold $\mathcal{M}_{\bf p}$ in (\ref{manifoldf}), the tangent space at any point ${\bf p} \in \mathbb{R}^L $ is isomorphic to $\mathbb{R}^L $ itself, given by,
\begin{equation}
 \mathrm{T}_{\bf p} \mathcal{M}_{\bf p} = \left\{ {\bm \xi}_{\mathbf{p}} \in \mathbb{R}^{L}  \right\}, \label{tsp}
\end{equation}
where ${\bm \xi}_{\mathbf{p}}$ is the tangent vector at the point ${\bf p}$. Since Euclidean space is flat, the tangent space at any point ${\bf p}$ consists of all possible vectors in $\mathbb{R}^L$, without any additional constraints. Geometrically, the tangent space at ${\bf p}$ is a linear space with the same dimension as the manifold, meaning it spans the entire $L$-dimensional space, representing all possible directions of movement within $\mathbb{R}^L$.

Based on (\ref{tsw}) and (\ref{tsp}), the UTS $\mathrm{T}_{({\bf w},{\bf p})} \mathcal{M}$ is expressed as \cite{zhong2024p},
\begin{equation}
    \begin{aligned}
        \mathrm{T}_{({\bf w},{\bf p})} \mathcal{M}& = \mathrm{T}_{\mathbf{w}} \mathcal{M}_{\bf w} \times \mathrm{T}_{\bf p} \mathcal{M}_{\bf p}\\&=\left\{\begin{array}{l}
        ({\bm \xi}_{\mathbf{w}}\in \mathbb{C}^{L},{\bm \xi}_{\mathbf{p}} \in \mathbb{R}^{L}): \\(\Re\left\{{\bm \xi}_{\mathbf{w}} \odot \mathbf{w}^*\right\}=\mathbf{0}_{L},{\bm \xi}_{\mathbf{p}})
        \end{array}\right\},
    \end{aligned}
    \label{UTS}
\end{equation}

\begin{figure}[!htbp]
  \begin{center}
  \includegraphics[width=3in]{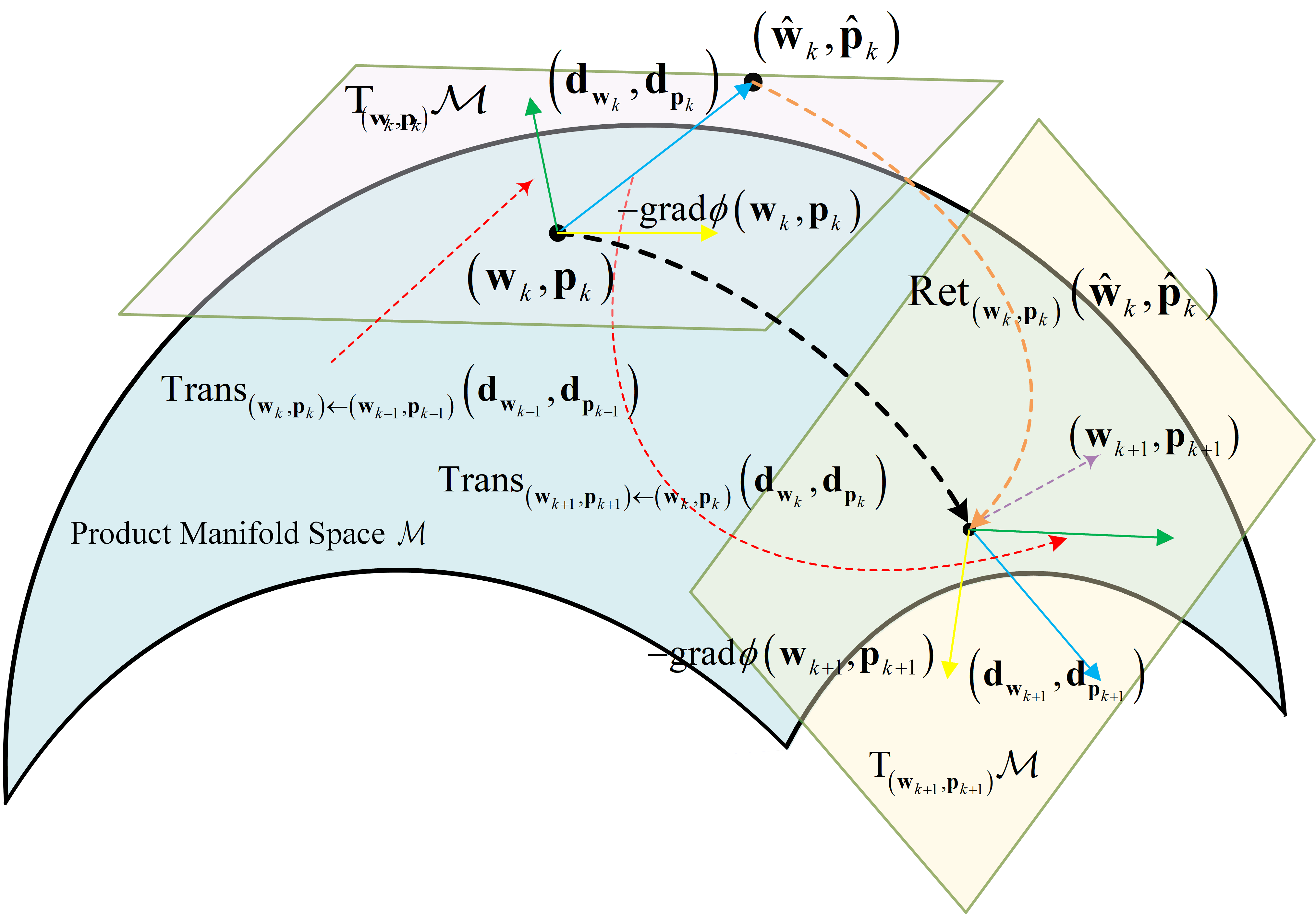}\\
  \caption{Geometric illustration of the PCPM method.}\label{PCPMframe}
  \end{center}
\end{figure}

By leveraging the conceptual foundations of PMS (\ref{productmanifold}) and UTS (\ref{UTS}), we systematically develop an efficient algorithm for solving (\ref{reforoverpro}), as depicted in Fig. \ref{PCPMframe}. The algorithm primarily employs CGD within the UTS, followed by projecting the solution back onto the UTS. A key aspect of this methodology is the strategic design of transportation operations that enable smooth transitions between different spaces within the manifold framework. This approach not only ensures computational efficiency but also highlights the importance of addressing the geometric complexities inherent in (\ref{reforoverpro}).

\subsubsection{The proposed CGD algorithm}
In this subsection, we propose the CGD algorithm to solve (\ref{reforoverpro}), where AB \(\bf w\) and MA positions \(\bf p\) are simultaneously optimized. Let \({\bf w}_0\) and \({\bf p}_0\) be the initial variables. As illustrated in Fig. \ref{PCPMframe}, the CGD algorithm cyclically performs the following three steps: (1) calculation of the parallel Riemannian gradient, (2) update of the feasible solution, and (3) adaptive adjustment of the step size using the Armijo line search, for \(k = 0, 1, \ldots\), until a stopping criterion is satisfied.

\textit{(a) Derive the parallel Riemannian gradient:} The parallel Riemannian gradient of $\phi({\bf w}_k, {\bf p}_k)$ in (\ref{reforoverpro}) is a combination of the individual Riemannian gradients with respect to ${\bf w}_k$ and ${\bf p}_k$. The Riemannian gradient is defined as the orthogonal projection of the Euclidean gradient onto the tangent space, which is expressed as \cite{tabrizi2025completederivationcomplexcircle},
\begin{subequations}
\begin{align}
 \operatorname{grad}_{{\bf w}_k} 
&=  \operatorname{Proj}_{{\bf w}_k}\left(\nabla_{{\bf w}_k} \phi({\bf w}_k, {\bf p}_k)\right) \notag\\&
=  \nabla_{{\bf w}_k} \phi({\bf w}_k, {\bf p}_k) -\Re\left[\nabla_{{\bf w}_k} \phi({\bf w}_k, {\bf p}_k) \odot {\bf w}_k^*\right] \odot {\bf w}_k   ,\\
\operatorname{grad}_{{\bf p}_k} &=  \operatorname{Proj}_{{\bf p}_k}\left(\nabla_{{\bf p}_k} \phi({\bf w}_k, {\bf p}_k)\right) = \nabla_{{\bf p}_k} \phi({\bf w}_k, {\bf p}_k),
\end{align} 
\label{rgfwp}%
\end{subequations}
where $\operatorname{grad}_{{\bf w}_k} $ and $\operatorname{grad}_{{\bf p}_k}$ are the orthogonal projection operators with respect to ${\bf w}_k$ and ${\bf p}_k$. $\nabla_{{\bf w}_k} \phi({\bf w}_k, {\bf p}_k)$ and $\nabla_{{\bf p}_k} \phi({\bf w}_k, {\bf p}_k)$ represent the Euclidean gradients with respect to ${\bf w}_k$ and ${\bf p}_k$. The derivations of the Euclidean gradients with respect to $\bf w$ and ${\bf p}$ are presented in Appendix \ref{appendixB}. 

Based on (\ref{rgfwp}), the parallel Riemannian gradient of $\phi({\bf w}_k, {\bf p}_k)$ on the UTS is given as,
\begin{equation}
    \operatorname{grad}\phi({\bf w}_k, {\bf p}_k) = [\operatorname{grad}_{{\bf w}_k} ;\operatorname{grad}_{{\bf p}_k}].\label{UTS2}
\end{equation}

\textit{(b) Update of the feasible solution:} Generally speaking, the current negative gradient is commonly used as the descent direction in manifold-based optimization methods. However, utilizing this descent direction often leads to getting stuck in local minima when the gradient approaches zero. To address this issue, we adopt a conjugate descent direction \cite{hager2006survey}, which considers not only the current gradient but also the gradients from previous iterations. By incorporating gradient information from prior iterations, this approach helps to escape local minima where pure gradient descent would otherwise become trapped. Additionally, the information from previous iterations accelerates convergence by amplifying movement in directions of consistent descent, enabling the algorithm to converge more efficiently. Specifically, the conjugate descent direction is given as follows,
\begin{equation}
    \begin{aligned}
        {\bf d}_{({\bf w}_k, {\bf p}_k)} &= [{\bf d}_{{\bf w}_k};{\bf d}_{{\bf p}_k}]\\
        &= \left[\begin{array}{l} -\operatorname{grad}_{{\bf w}_k}   +\sigma_k \operatorname{Tran}_{{\bf w}_k \leftarrow {\bf w}_{k-1} } ({\bf d}_{{\bf w}_{k-1}});\\- \operatorname{grad}_{{\bf p}_k}   +\sigma_k \operatorname{Tran}_{{\bf p}_k \leftarrow {\bf p}_{k-1} } ({\bf d}_{{\bf p}_{k-1}}) \end{array}\right]
    \end{aligned}
    \label{ddt}
\end{equation}
where ${\bf d}_{{\bf w}_k}$ and ${\bf d}_{{\bf p}_k}$ represent the individual descent directions for ${\bf w}_k$ and ${\bf p}_k$, respectively; $\sigma_k$ is the conjugate parameter which is deﬁned in (\ref{longlongterm}) at the top of the next page; $\operatorname{Tran}_{{\bf w}_k \leftarrow {\bf w}_{k-1} }({\bf d}_{{\bf w}_{k-1}})$ and $\operatorname{Tran}_{{\bf p}_k \leftarrow {\bf p}_{k-1} }({\bf d}_{{\bf p}_{k-1}})$ represent the transport operations that map tangent vectors between different tangent spaces. These operations transfer the tangent vectors ${\bf d}_{{\bf w}_{k-1}}$ and ${\bf d}_{{\bf p}_{k-1}}$ from the points ${\bf w}_{k-1}\in\mathcal{M}$ and ${\bf p}_{k-1}\in\mathcal{M}$ to new points ${\bf w}_k\in\mathcal{M}$ and ${\bf p}_k\in\mathcal{M}$, respectively, as follows,
\begin{figure*}[!t]
	\begin{align}
		\sigma_k=\left[ \frac{\Re\left[\begin{array}{c}
\operatorname{grad}_{{\bf w}_k}^H\left(\operatorname{grad}_{{\bf w}_k}-\left(\operatorname{grad}_{{\bf w }_{k-1}}-\Re\left\{\operatorname{grad}_{{\bf w }_{k-1}} \odot {\bf w}_k^*\right\} \odot {\bf w}_k\right)\right) 
+\operatorname{grad}_{{\bf p}_k}^H\left(\operatorname{grad}_{{\bf p}_k}-\operatorname{grad}_{{\bf p}_{k-1}}\right)
\end{array}\right] }{\Re\left[\begin{array}{l}
{\left(     \operatorname{grad}_{{\bf w}_k}    -\left(\operatorname{grad}_{{\bf w }_{k-1}}-\Re\left\{\operatorname{grad}_{{\bf w }_{k-1}} \odot {\bf w}_k^*\right\} \odot {\bf w}_k\right) \right)^H 
 \left({\bf d}_{{\bf w}_{k-1}}-\Re\left\{{\bf d}_{{\bf w}_{k-1}} \odot {\bf w}_k^*\right\} \odot {\bf w}_k \right)} \\\quad\quad\quad\quad\quad\quad\quad\quad\quad\quad\quad\quad+\left(\operatorname{grad}_{{\bf p}_k}-\operatorname{grad}_{{\bf p}_{k-1}}\right)^H {\bf d}_{{\bf p}_{k-1}}
\end{array} \right]} \right]^+
\label{longlongterm}	
\end{align}
 \hrulefill
\end{figure*}
\begin{subequations}
    \begin{align}
       & \operatorname{Tran}_{{\bf w}_k \leftarrow {\bf w}_{k-1} }({\bf d}_{{\bf w}_{k-1}}) =  {\bf d}_{{\bf w}_{k-1}} -   \Re\left[{\bf d}_{{\bf w}_{k-1}} \odot {\bf w}_k^*\right] \odot {\bf w}_k,\\
       & \operatorname{Tran}_{{\bf p}_k \leftarrow {\bf p}_{k-1} }({\bf d}_{{\bf p}_{k-1}}) =  {\bf d}_{{\bf p}_{k-1}}.
    \end{align}
    \label{transd}
\end{subequations}

Given the descent direction ${\bf d}_{({\bf w}_k, {\bf p}_k)}$ in (\ref{ddt}), the update on the UTS $\mathrm{T}_{({\bf w}_k,{\bf p}_k)} \mathcal{M}$ is,
\begin{equation}
\begin{aligned}
    [{\bf{\hat w}}_k;{\bf{\hat p}}_k]&=[{\bf{ w}}_k;{\bf{ p}}_k] + \upsilon_k  {\bf d}_{({\bf w}_k, {\bf p}_k)} \\ &= [{\bf{ w}}_k+\upsilon_k{\bf d}_{{{\bf w}}_k};{\bf{ p}}_k+\upsilon_k{\bf d}_{{{\bf p}}_k}],
\end{aligned}
\label{uUTSTM}
\end{equation}
where $\upsilon_k$ is the step size, adaptively updated by the Armijo line search algorithm, as described in the next step; ${\bf{\hat w}}_k$ and ${\bf{\hat p}}_k$ are the updated values on the UTS. Generally, updates on the UTS $\mathrm{T}_{({\bf w}_k,{\bf p}_k)} \mathcal{M}$ may not ensure that the resulting point remains on the PMS $\mathcal{M}$. In such cases, a retraction process is necessary to map a point from the UTS $\mathrm{T}_{({\bf w}_k,{\bf p}_k)} \mathcal{M}$ back onto the PMS $\mathcal{M}$, thereby ensuring feasibility. Hence, the next feasible solution is obtained as follows,
\begin{equation}
\begin{aligned}
    [{\bf  w}_{k+1};{\bf  p}_{k+1}]&= \text{Ret}_{({\bf{ w}}_k,{\bf{ p}}_k)}({\bf{\hat w}}_k,{\bf{\hat p}}_k) \\&= [\text{Ret}_{{\bf{ w}}_k}({\bf{\hat w}}_k);\text{Ret}_{{\bf{ p}}_k}({\bf{\hat p}}_k)]\\&=\left[{\bf{\hat w}}_k \odot \frac{1}{|{\bf{\hat w}}_k|}; {\bf{\hat p}}_k  \right],
\end{aligned}
\label{recf}
\end{equation}
where $\text{Ret}_{{\bf{ w}}_k}({\bf{\hat w}}_k)$ and $\text{Ret}_{{\bf{ p}}_k}({\bf{\hat p}}_k)$ denote the individual retraction operations for ${\bf{\hat w}}_k$ and ${\bf{\hat p}}_k$, respectively.

\textit{(c) Adaptive step size update:} Using the Armijo line search strategy \cite{dai2002conjugate}, we dynamically adjust the step size during updates, allowing the algorithm to converge more efficiently. By modifying the step size based on the improvement in the objective function, the algorithm takes larger steps when progress is significant and smaller steps when progress slows. Specifically, the Armijo line search strategy is defined as follows,
\begin{equation}
  \phi({\bf w}_{k+1}, {\bf p}_{k+1})  \le \phi({\bf w}_k, {\bf p}_k) + \tau^r \cdot {\hat \upsilon} \cdot {\hat d}_{\phi_k}, \label{Als}
\end{equation}
where $\tau \in (0,1)$ is the step size coefficient, \(r\) is the number of line searches, and \({\hat \upsilon}\) is the initial step size. The final step size for the \(k\)-th iteration is given by \(\upsilon_{k}=\tau^r \cdot  {\hat \upsilon} \). \({\hat d}_{\phi_k}\) represents the directional derivative, which is defined as follows:
\begin{equation}
{\hat d}_{\phi_k}=\operatorname{grad}^H\phi({\bf w}_k, {\bf p}_k)  {\bf d}_{({\bf w}_k, {\bf p}_k)}.     \label{Als2}
\end{equation}
To further expedite convergence, the initial step size ${\hat \upsilon}$ for the upcoming $(k+1)$-th iteration can be adjusted based on changes in $r$. If $r=1$ satisﬁes (\ref{Als}), it means that only one line search was performed, resulting in a decrease in the objective function. This indicates that the current step size is too small, so it should be increased to ${\hat \upsilon}=\iota_1\upsilon_k$ for the next iteration, where $\iota_1>1$. If $r=2$ satisﬁes (\ref{Als}), it indicates an appropriately chosen initial step size, necessitating that the step size remains at ${\hat \upsilon}=\upsilon_k$. If $r\ge 3$ satisﬁes (\ref{Als}), it indicates that more than three line searches were performed, resulting in a small step size. In this case, the step size should be increased to ${\hat \upsilon}=\iota_2\upsilon_k$ for the next iteration, where $\iota_2>1$. In summary, the initial step size ${\hat \upsilon}$ update rule for the next iteration is summarized as,
\begin{equation}
{\hat \upsilon} = \left\{ {\begin{array}{*{20}{c}}
{\iota_1\upsilon_k},r=1\\
{\upsilon_k},r=2\\
{\iota_2\upsilon_k},r>2
\end{array}} \right ..
\end{equation}

Based on the above discussion, the proposed CGD algorithm to tackle (\ref{reforoverpro}) is compactly presented in Algorithm \ref{alg:1}. The algorithm mainly includes: (1) calculation of the parallel Riemannian gradient, (2) update of the feasible solution, and (3) adaptive adjustment of the step size using the Armijo line search. The stop conditions include: the Riemannian gradient satisﬁes $\|\operatorname{grad}\phi({\bf w}_k, {\bf p}_k)\|\le\varepsilon_j$ or the number of iterations satisﬁes $k\ge200$.
\begin{algorithm}
	\floatname{algorithm}{Algorithm}
	\renewcommand{\algorithmicrequire}{\textbf{Input:}}
	\renewcommand{\algorithmicensure}{\textbf{Output:}}
	\caption{: The CGD algorithm to the problem (\ref{reforoverpro}).}
	\label{alg:1}
	\begin{algorithmic}[1]
		\REQUIRE 
                ${\bf w}_j, {\bf p}_j,\varepsilon_{j} $.\\
                \ENSURE
                ${\bf w}_{j+1}={\bf w}_{k+1}, {\bf p}_{j+1}={\bf p}_{k+1}$.\\
            \STATE Initialize $k = 0$, ${\bf w}_k={\bf w}_j$ and ${\bf p}_k={\bf p}_j$.   
            \STATE {\textbf {repeat}}
            \STATE \quad Calculate $\operatorname{grad}\phi({\bf w}_k, {\bf p}_k)$ by (\ref{rgfwp})-(\ref{UTS2});
            \STATE \quad Calculate ${\bf d}_{({\bf w}_k, {\bf p}_k)}$ by (\ref{ddt})-(\ref{transd}) and $\upsilon_k$ by (\ref{Als})-(\ref{Als2});
            \STATE \quad Update ${\bf w}_{k+1}$ and ${\bf p}_{k+1}$ by (\ref{uUTSTM})-(\ref{recf});
            \STATE \quad $k \leftarrow k+1$;
            \STATE {\textbf {end}}
		\STATE {\textbf {Until} some stopping criterions (e.g., $\|\operatorname{grad}\phi({\bf w}_k, {\bf p}_k)\|\le\varepsilon_j$ or $k\ge200$}) are satisfied\\
	\end{algorithmic}%
\end{algorithm}

\subsubsection{Update the smoothing parameter, the penalty parameter, threshold parameter, and constraint violation parameter}

The smoothing parameter $\gamma$ and the penalty parameter \(\rho\) play a crucial role in balancing constraints and solutions in optimization problems, allowing for control over solution properties and the optimization process. In this work, we gradually decrease $\gamma$ in each iteration of the outer loop according to, $\gamma_{j+1}=\max\{\gamma_{\min},\gamma_{j}\cdot\delta_{\gamma}\}$, where $\gamma_{\min}$ is the minimum allowable value, and $\delta_{\gamma} \in (0,1)$ is a constant decay factor. Meanwhile, the penalty parameter \(\rho\) is gradually increased during each outer iteration. Specifically, if the current solution is far from satisfying the constraints, it may indicate that the penalty is insufficient. In such cases, $\rho$ is updated for the next iteration as \(\rho_{j+1} = \rho_j / \delta_{\rho}\), where \(\delta_{\rho} \in (0, 1)\) is a constant factor that adjusts the penalty parameter.  

The threshold parameter \(\varepsilon_j\) is used to limit the Riemannian gradient Frobenius norm of the CGD algorithm, represented as \(\|\operatorname{grad}\phi({\bf w}_k, {\bf p}_k)\| \le \varepsilon_j\). As iterations progress, more accurate solutions are required, necessitating a reduction in \(\varepsilon_j\). Therefore, the adjustment is made as \(\varepsilon_{j+1} = \max\{\varepsilon_{\min}, \varepsilon_j \cdot \delta_{\varepsilon}\}\), where \(\varepsilon_{\min}\) is the lower bound, and \(\delta_{\varepsilon} \in (0, 1)\) is a constant factor for adjusting \(\varepsilon_j\).

The constraint violation parameter \(\varsigma_j\) is used to constrain the spectral inequality \(g_l(p_{l+1}, p_{l}), \forall l\), \(f_1(p_{1})\), and \(f_2(p_{L})\) in (\ref{totalett}) to satisfy \(\max\{0, g_l(p_{l+1}, p_{l}), \forall l, f_1(p_{1}), f_2(p_{L})\} \le \varsigma_j \to \varsigma_{\min} = 0\), where \(\varsigma_{\min}\) is the lower bound of \(\varsigma_j\). As iterations progress, the CGD algorithm generates sequences \({\bf w}_j\) and \({\bf p}_j\) that converge towards a feasible limit point satisfying \(g_l(p_{l+1}, p_{l}) \le 0, \forall l\), \(f_1(p_{1}) \le 0\), and \(f_2(p_{L}) \le 0\) when the penalty factor \(\rho_j\) is sufficiently large. Therefore, the adjustment is made as \(\varsigma_{j+1} = \max\{\varsigma_{\min}, \varsigma_j \cdot \delta_{\varsigma}\}\), where $\varsigma_{\min}$ is the lower bound, and \(\delta_{\varsigma} \in (0, 1)\) is a constant factor for adjusting \(\varsigma_j\).

To obtain a near-optimal solution for (\ref{reforset}), the PCPM framework is detailed in Algorithm \ref{alg:2}. The termination conditions, denoted by \((\gamma_{\min},o_{\min}, \varepsilon_{\min}, \varsigma_{\min})\), serve as essential lower bounds within the proposed algorithm, ensuring convergence and computational efficiency.
\begin{algorithm}
	\floatname{algorithm}{Algorithm}
	\renewcommand{\algorithmicrequire}{\textbf{Input:}}
	\renewcommand{\algorithmicensure}{\textbf{Output:}}
	\caption{: The PCPM framework to the problem (\ref{reforset}).}
	\label{alg:2}
	\begin{algorithmic}[1]
		\REQUIRE 
                Initial $j=0$, CM beamforming ${\bf w}_j$, position of MAs ${\bf p}_j$, smoothing parameter $\gamma_j$, minimum smoothing parameter $\gamma_{\min}$, penalty parameter ${\rho}_j$, threshold parameter $\varepsilon_j$, minimum threshold parameter $\varepsilon_{\min}$, constraint violation parameter $\varsigma_j$, minimum constraint violation parameter $\varsigma_{\min}$, constant factors $\delta_{\gamma},\delta_{\rho},\delta_{\varepsilon},\delta_{\varsigma} \in (0, 1)$, and minimum step length $o_{\min}$.\\
            \STATE {\textbf {Reapeat}}
            \STATE \quad Update ${\bf w}_{j+1}$ and ${\bf p}_{j+1}$ by \textbf{Algorithm 1};
            \STATE \quad Update $\gamma_{j+1} = \max\{\gamma_{\min}, \gamma_j \cdot \delta_{\gamma}\}$;
            \STATE \quad Update $\varepsilon_{j+1} = \max\{\varepsilon_{\min}, \varepsilon_j \cdot \delta_{\varepsilon}\}$; 
            \STATE \quad Update $\varsigma_{j+1} = \max\{\varsigma_{\min}, \varsigma_j \cdot \delta_{\varsigma}\}$;
            \STATE \quad \textbf{IF} $\max\left\{\begin{array}{l} \max\{0,g_l(p_{l+1},p_{l})\},\forall l, \\ \max\{0,f_1(p_{1})\},\max\{0,f_2(p_{L})\}  \end{array} \right\} \ge\varsigma_{j+1}$
            \STATE \quad\quad Update $\rho_{j+1} = \rho_j / \delta_{\rho}$;
            \STATE \quad \textbf{ELSE}
            \STATE \quad\quad Update $\rho_{j+1} = \rho_j$;
            \STATE \quad \textbf{END}
            \STATE \quad$j \leftarrow j+1$;
		\STATE {\textbf {Until} some stopping criterions (e.g., $\| [{\bf w}_{j};{\bf p}_{j}]-[{\bf w}_{j-1};{\bf p}_{j-1}]\|\le o_{\min}$, $\gamma_j \le \gamma_{\min}$, $\varepsilon_j \le \varepsilon_{\min}$, and $\varsigma_{j} \le \varsigma_{\min}$ }) are satisfied\\
	\end{algorithmic}%
\end{algorithm}

\subsection{Analysis of computation complexity and convergence}
\subsubsection{Analysis of computation complexity} The computational complexity of the proposed PCPM framework primarily arises from updating AB vector ${\bf w}$ and MA positions ${\bf p}$ in Algorithm \ref{alg:1}. Specifically, computing the Riemannian gradients $\operatorname{grad}{\bf w}_k$ and $\operatorname{grad}{\bf p}_k$ with respect to ${\bf w}$ and ${\bf p}$, as shown in equations (\ref{rgfwp}) and (\ref{UTS2}), has approximate computational complexities of ${\cal O}(M_p(N_e+N_b)(2L^2+2L))$ and ${\cal O}(M_p(N_e+N_b)(12L^2+4L))$, respectively. The computation of the descent direction ${\bf d}_{({\bf w}_k, {\bf p}_k)}$ according to equations (\ref{ddt})-(\ref{transd}) has an approximate complexity of ${\cal O}(4L)$. Additionally, determining the step size $\upsilon_k$ via equations (\ref{Als}) and (\ref{Als2}) requires approximately ${\cal O}(2rL)$ complexity. Assuming $T_{in}$ iterations for the inner loop in Algorithm \ref{alg:1} and $T_{out}$ iterations for the outer loop in Algorithm \ref{alg:2}, the overall complexity of the PCPM framework is approximately ${\cal O}((T_{in}+T_{out})M_p(N_e+N_b)(14L^2+(10+2r)L))$.

\subsubsection{Analysis of convergence} To establish the convergence for Algorithm \ref{alg:2}, we first prove the convergence of Algorithm \ref{alg:1} to a stationary point in Theorem \ref{theorem1}.

\begin{theorem} \label{theorem1} \itshape
Let $\{{\bf w},{\bf p}\}$ be a sequence generated by Algorithm \ref{alg:1}. Then, every limit point generated by this sequence is a stationary point of problem (\ref{reforoverpro}).
\end{theorem}

\quad {\bf\textit{Proof}}{\bf:} See Appendix \ref{appendixC}. $\hfill\blacksquare$

We now demonstrate that Algorithm \ref{alg:2} converges to a KKT point. To this end, we utilize the following theorem from \cite{liu2020simple}.

\begin{theorem} \label{theorem2} \itshape
In Algorithm \ref{alg:2}, we set $\gamma_{\min}=o_{\min}= \varepsilon_{\min}= \varsigma_{\min}=0$ and $\rho_0$ (the initial penalty) is
sufficiently large. If the sequence $\{{\bf w},{\bf p}\}$ produced by Algorithm \ref{alg:1} admits a feasible stationary point where the linear independence constraint qualification (LICQ) condition holds, then it also satisfies the KKT conditions for (\ref{reforset}). \end{theorem}

\quad {\bf\textit{Proof}}{\bf:} See \textit{Proposition 4.2} in \cite{liu2020simple}. $\hfill\blacksquare$

To ensure that Algorithm \ref{alg:2} converges to a KKT point, it remains to verify that the sequence $\{{\bf w},{\bf p}\}$ satisfies the LICQ condition \cite{wachsmuth2013licq}. We analyze the gradients of all active constraints in (\ref{reforset}) at any candidate point,
\begin{itemize}
    \item Equality constraints $|w_l| = 1, \, l = 1, \dots, L$ in (\ref{CMC}) have gradients $\nabla h_{w_l}$ aligned with the canonical basis vector corresponding to $w_l$. Since each constraint involves a distinct component of ${\bf w}$, these gradients are linearly independent.
    \item Minimum spacing constraints $p_{l+1} - p_l \geq \lambda/2$ (if active) in (\ref{minidistance}) have gradients of the form $\nabla h_{p_l} = [0, \dots, -1, 1, \dots, 0]$, which involve only two consecutive position variables and are mutually independent.
    \item Boundary constraints $p_1 \geq 0$ and $p_L \leq D$ in (\ref{antennaselect}) have gradients $\nabla h_{p_1} = [1, 0, \dots, 0]$ and $\nabla h_{p_L} = [0, \dots, 0, 1]$, which are distinct canonical basis vectors in the position space.
\end{itemize}
Since the gradients of all active constraints are mutually independent, the LICQ condition holds for problem (\ref{reforset}). Therefore, with $\gamma_{\min}, o_{\min}, \varepsilon_{\min}, \varsigma_{\min} \to 0$ and a sufficiently large $\rho_0$, Algorithm \ref{alg:2} converges to a KKT point by Theorem \ref{theorem2}.

\section{Numerical Results}
In this section, we present simulation results to evaluate the PLS performance of the MA-enabled design in the proposed MISO communication system with AB. For comparison, the proposed method is evaluated against the following benchmark schemes:
\begin{itemize}
    \item \textbf{MA-FDB-GD}: The transmitter is equipped with an FDB architecture and \( L \) MAs, where the gradient descent method proposed in \cite{hu2024secure,cheng2024enabling} is applied.
    \item \textbf{FPA-FDB-SS}: The transmitter is equipped with an FDB architecture and FPA-based uniform linear arrays (ULAs), where \( L \) transmit antennas are selected via greedy search \cite{bose2021efficient} to maximize the MSR.
    \item \textbf{MA-AB-GD}: Similar to the MA-FDB-GD architecture, but it incorporates a direct projection operation to ensure that the FDB vector lies on the unit circle, achieving CM AB.
    \item \textbf{FPA-FDB-ULA}: The transmitter is equipped with an FDB architecture and \( L \) FPA-based ULAs \cite{li2019constant}.
    \item \textbf{FPA-AB-ULA}: The transmitter is equipped with an AB architecture and \( L \) FPA-based ULAs \cite{li2019constant}.
    \item \textbf{MA-AB-R}: The transmitter is equipped with an AB architecture and \( L \) MAs, where the MAs are distributed randomly on average to satisfy the constraints of the MA movement area.
\end{itemize}

In the simulations, the parameters are set as follows, unless otherwise specified: the number of transmit antennas is \( L = 16 \), the wavelength is \( \lambda = 0.01 \, \text{m} \), the total transmitter power is \( P_t = MP = 0 \, \text{dBW} \), and the total length of the antenna array is \( D = 30\lambda \). The number of LUs is assumed to be \( N_b = 4 \), and the number of EVEs is \( N_e = 4 \). Additionally, the distance between each $b$-th LU or $e$-th EVE and the BS is modeled as a random variable following a uniform distribution, i.e., $d_i \sim \mathcal{U}[d_{\min}, d_{\max}], \forall i\in\{b,e\}$, where $d_{\min} = 60\text{m}$ and $d_{\max} = 100\text{m}$ represent the minimum and maximum distances, respectively. The spatial AoDs from the BS to the $b$-th LU and the $e$-th EVE are assumed to be independent and identically distributed (i.i.d.) random variables drawn from a uniform distribution, i.e., ${\hat \theta_{b,m}}, {\check \theta_{e,m}} \sim \mathcal{U}[0,\pi]$, for $1\le m\le M_p$. The total number of propagation paths is set as $M_p = 6$. Correspondingly, the complex path gains for each $b$-th LU and $e$-th EVE are given by $\hat{\beta}_{b,m}\sim {g_0 d_b^{-\alpha}}/{M_p}$ and $\check{\beta}_{e,m}\sim {g_0 d_e^{-\alpha_{pl}}}/{M_p}$, respectively, for $1 \le m \le M_p$, where $g_0 = -40\text{dB}$ denotes the average channel gain at the reference distance $d_0 = 1\text{m}$, and $\alpha_{pl} = 2.8$ represents the path-loss exponent. Furthermore, the variances of Gaussian noise are set as ${\hat\sigma}_b^2 = {\check\sigma}_e^2 = -70\text{dBm}$, and the hyperparameter is set to $\alpha = 10^{0}$. All simulation results are averaged over 1,000 independent random realizations.

\begin{figure}[t]
  \begin{center}
  \includegraphics[width=3in]{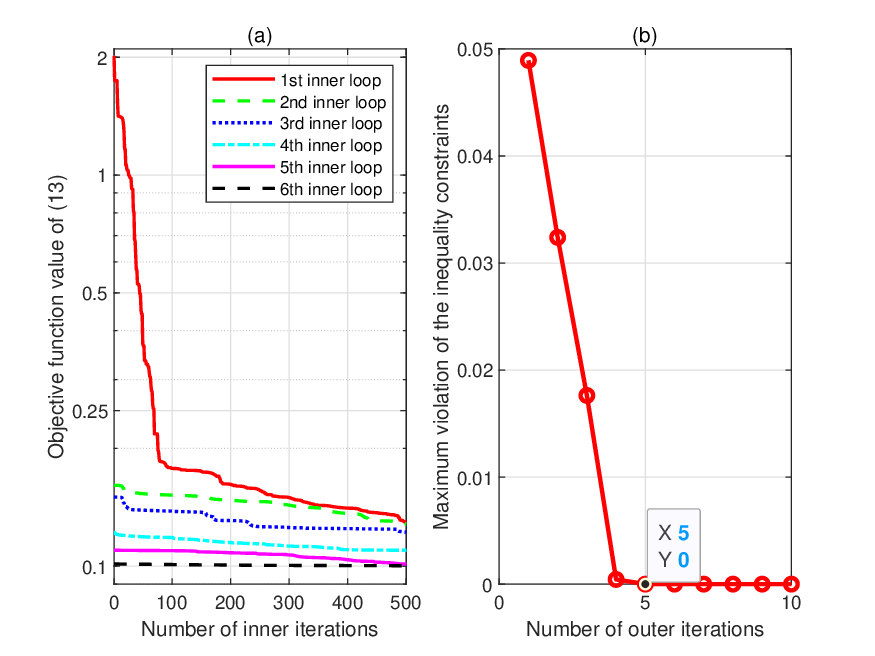}\\
  \caption{Convergence of the proposed PCPM framework.}\label{converg}
  \end{center}
\end{figure}

Fig. \ref{converg} illustrates the convergence performance of the proposed PCPM framework. In Fig. \ref{converg}(a), the objective function value in (\ref{reforset}) converges monotonically within each inner loop, with the required iterations decreasing significantly after the initial loop. The outer loop also converges monotonically, as the objective value at the end of each iteration is consistently lower than at the end of the previous one. Fig. \ref{converg}(b) shows the maximum violation of the inequality constraints, where the error converges rapidly within five outer iterations, confirming the fast convergence of the proposed framework.

\begin{figure}[t]
  \begin{center}
  \includegraphics[width=3in]{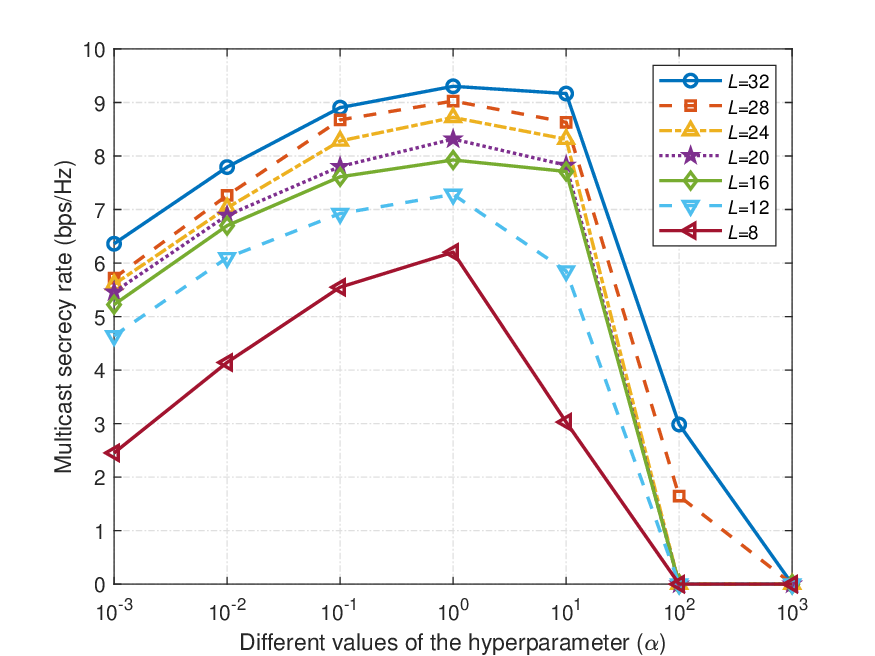}\\
  \caption{Comparison of MSR with varying values of the hyperparameter ($\alpha$).}\label{alpha}
  \end{center}
\end{figure}

Since the choice of $\alpha$ introduced in (\ref{reforset}) greatly influences the tightness of the upper bound for the original function, we vary its value in Fig. \ref{alpha} to determine the setting that maximizes PLS performance. As shown in the figure, for all considered numbers of transmit antennas $L$, the MSR increases monotonically as $\alpha$ grows from $10^{-3}$ to $10^{0}$, and then decreases monotonically as $\alpha$ increases from $10^{0}$ to $10^{3}$. The maximum MSR is achieved around $\alpha = 10^{0}$, indicating that this value provides the best approximation to the original function. Therefore, $\alpha = 10^{0}$ is selected as the hyperparameter in our design.

\begin{figure}[t]
  \begin{center}
  \includegraphics[width=3in]{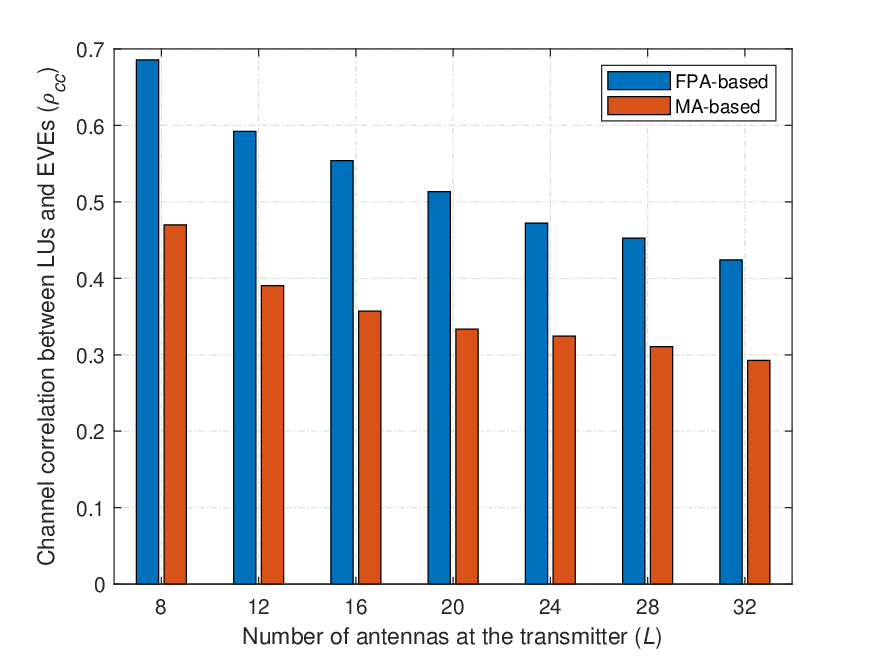}\\
  \caption{Comparison of channel correlations ($\rho_{cc}$) with varying numbers of antennas at the transmitter ($L$).}\label{cc}
  \end{center}
\end{figure}

Fig. \ref{cc} compares the channel correlation $\rho_{cc}$ between LUs and EVEs for the FPA-based and MA-based designs. The channel correlation is defined as $\rho_{cc}={\max }_{ b=1,...,N_b \atop e=1,...,N_e}\frac{|{\bf g}_e^H{\bf h}_b|}{\|{\bf g}_e\|_2\|{\bf h}_b\|_2}$. It can be observed that, with the assistance of MAs, $\rho_{cc}$ between LUs and EVEs is significantly reduced compared with the FPA-based systems. This reduction provides MA-based systems with greater spatial diversity, thereby enhancing PLS. Furthermore, both FPA-based and MA-based systems achieve lower $\rho_{cc}$ as the number of transmit antennas increases, indicating that increasing antennas is an effective way to further improve PLS. This observation underscores the importance of incorporating AB into MA-based systems to realize practical, low-cost designs. This observation also implies that in scenarios with high user density, where legitimate and eavesdropping channels are inherently strongly correlated, achieving a non-zero secrecy rate can be challenging. While MAs significantly reduce this correlation, as shown in the figure, they cannot completely eliminate it if an EVE is physically co-located with an LU. In such worst-case zero-secrecy scenarios, the channel vectors become nearly identical ($\rho_{cc}\to1$), rendering spatial separation infeasible.

\begin{figure}[t]
  \begin{center}
  \includegraphics[width=3in]{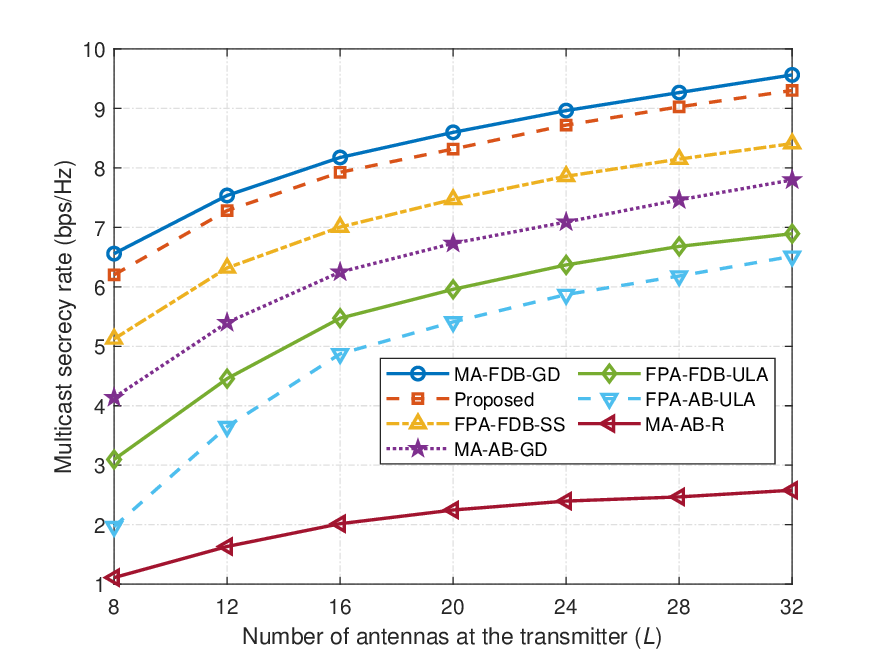}\\
  \caption{Comparison of MSR with varying numbers of antennas at the transmitter ($L$).}\label{da}
  \end{center}
\end{figure}

Fig. \ref{da} compares the MSR in (\ref{MCSR}) across various benchmark schemes as the number of antennas \( L \) increases. The results show that the MSR improves with more antennas due to the additional DoFs. The proposed method achieves the second-highest MSR, trailing only the MA-FDB-GD scheme, but significantly reduces hardware costs,making the proposed strategy more practical. It also outperforms all architectures without MAs, demonstrating the advantages of MAs in enhancing secure communication by modifying steering vectors. In contrast, the MA-AB-GD method underperforms compared to the FPA-FDB-SS scheme with sparse antenna selection, highlighting the limitations of the gradient descent approach for AB systems and the effectiveness of the proposed method.

\begin{figure}[t]
  \begin{center}
  \includegraphics[width=3in]{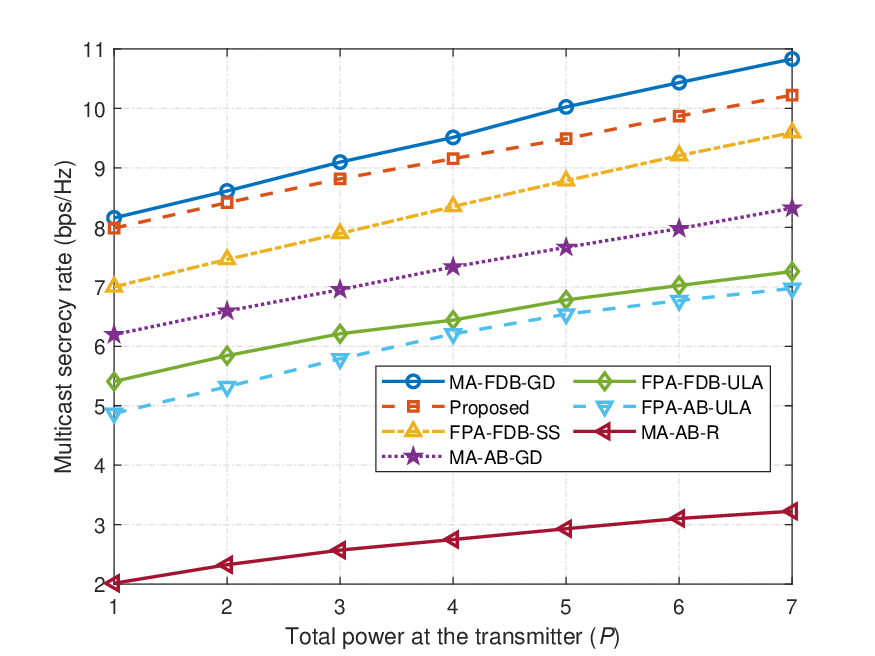}\\
  \caption{Comparison of MSR with varying total transmit power ($P_t$).}\label{dp}
  \end{center}
\end{figure}

Fig. \ref{dp} compares the MSR across different benchmark schemes as the total power at the transmitter ($P_t$) increases. As shown in the figure, the MSR increases with $P_t$ for all architectures, as higher power strengthens the signal in the system. However, the MA-AB-R method shows the slowest increase in MSR, which can be attributed to the random distribution of the antenna array. Unlike structured antenna arrays, the random positioning of antennas in the MA-AB-R method does not provide an optimal configuration for enhancing signal strength or spatial diversity, leading to less efficient utilization of the increased transmit power.

\begin{figure}[t]
  \begin{center}
  \includegraphics[width=3in]{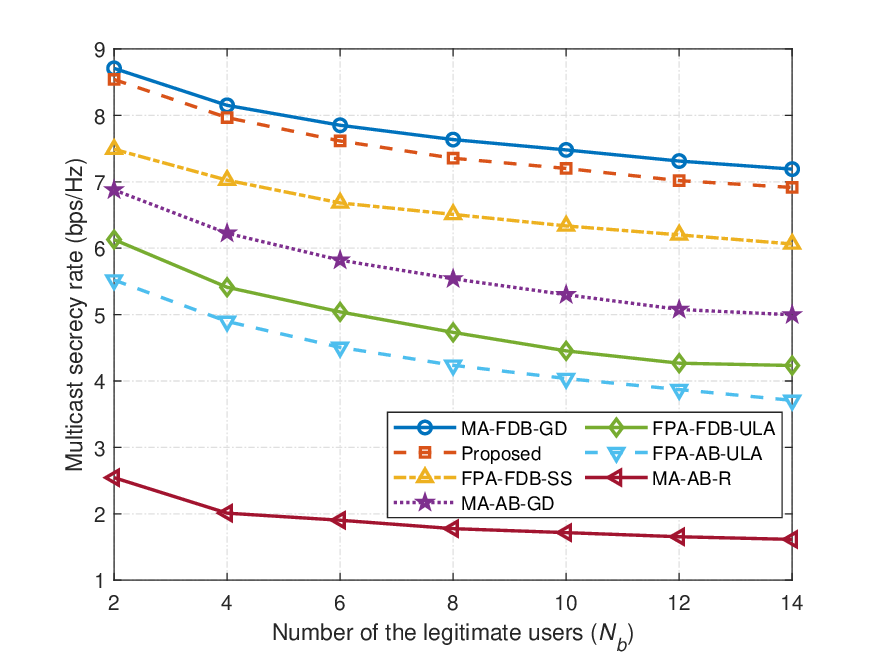}\\
  \caption{Comparison of MSR with varying numbers of legitimate users ($N_b$).}\label{db}
  \end{center}
\end{figure}

\begin{figure}[t]
  \begin{center}
  \includegraphics[width=3in]{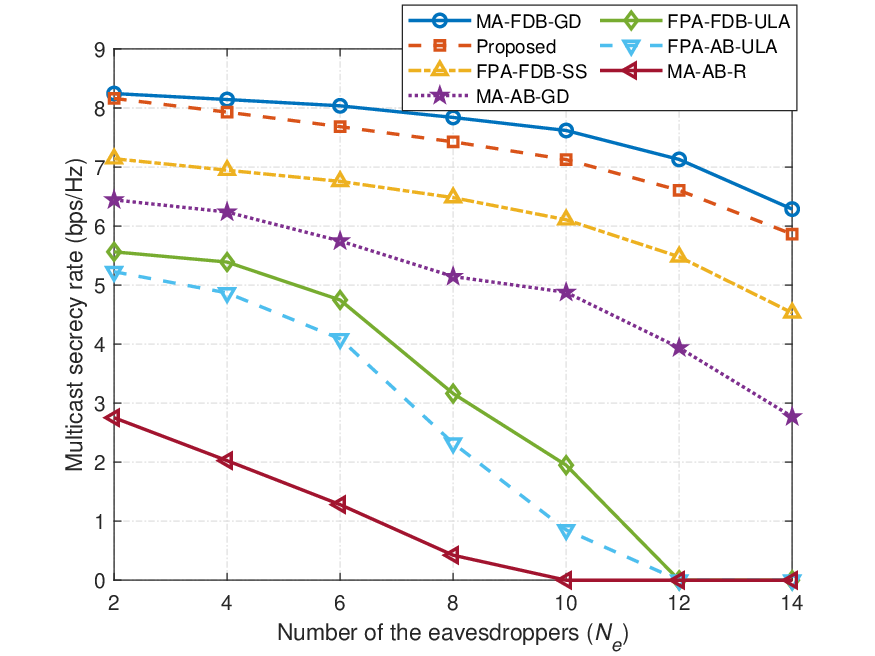}\\
  \caption{Comparison of MSR with varying numbers of eavesdroppers ($N_e$).}\label{de}
  \end{center}
\end{figure}

Figs. \ref{db} and \ref{de} depict the MSR performance of various architectures as the number of LUs ($N_b$) and EVEs ($N_e$) increases from 2 to 14. In both figures, the MSR decreases as $N_b$ and $N_e$ increase, primarily due to heightened interference and signal leakage, which make maintaining secrecy more challenging. Notably, the rate of decline is more pronounced with increasing EVEs compared to LUs. This is because a larger number of EVEs amplifies signal leakage to unintended receivers, significantly reducing the communication secrecy. This trend highlights a critical operational limit: as $N_e$ continues to grow, the probability of at least one EVE possessing a channel stronger than that of the weakest LU increases. Eventually, the MSR for all schemes will inevitably converge to zero, defining a zero-secrecy scenario. Our proposed method, by effectively leveraging MAs to mitigate channel correlation, can sustain a positive secrecy rate for a larger number of EVEs compared to FPA-based schemes, thereby pushing back the onset of this zero-secrecy threshold. In contrast, while additional LUs introduce interference, their impact is less severe as they are part of the intended communication. Despite these challenges, the proposed method consistently achieves the second-highest MSR in both scenarios, underscoring its superior performance and demonstrating the effectiveness of MA-enabled systems with hardware-efficient AB in optimizing antenna positions for secure multicast transmission in dynamic environments.

\begin{figure}[t]
  \begin{center}
  \includegraphics[width=3in]{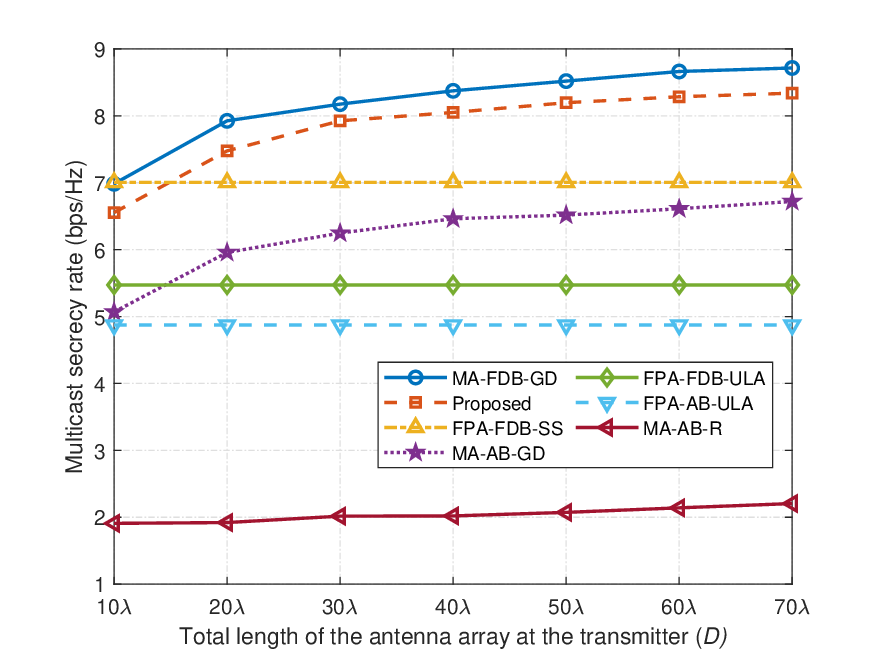}\\
  \caption{Comparison of MSR with varying total antenna array length at the transmitter ($D$).}\label{dl}
  \end{center}
\end{figure}

Fig. \ref{dl} shows the MSR performance of various architectures as the total antenna array length \( D \) increases from \( 10\lambda \) to \( 70\lambda \). The results indicate that the MSR of MA-enabled systems increases with \( D \), while FPA-based systems remain nearly constant. This is because a larger \( D \) provides more spatial DoFs for MAs to dynamically adjust antenna positions, optimizing beamforming toward legitimate users and improving multicast secrecy. In contrast, FPA systems with ULAs, having fixed antenna positions, lack adaptability and do not benefit from an increase in \( D \). Notably, the MSR of MA-FDB-GD and the proposed method starts lower at \( D = 10\lambda \) compared to FPA-FDB-SS but surpasses it as \( D \) increases. This demonstrates the superior capability of MA-enabled systems to exploit larger array configurations for enhanced secrecy performance.

\begin{figure}[t]
  \begin{center}
  \includegraphics[width=3in]{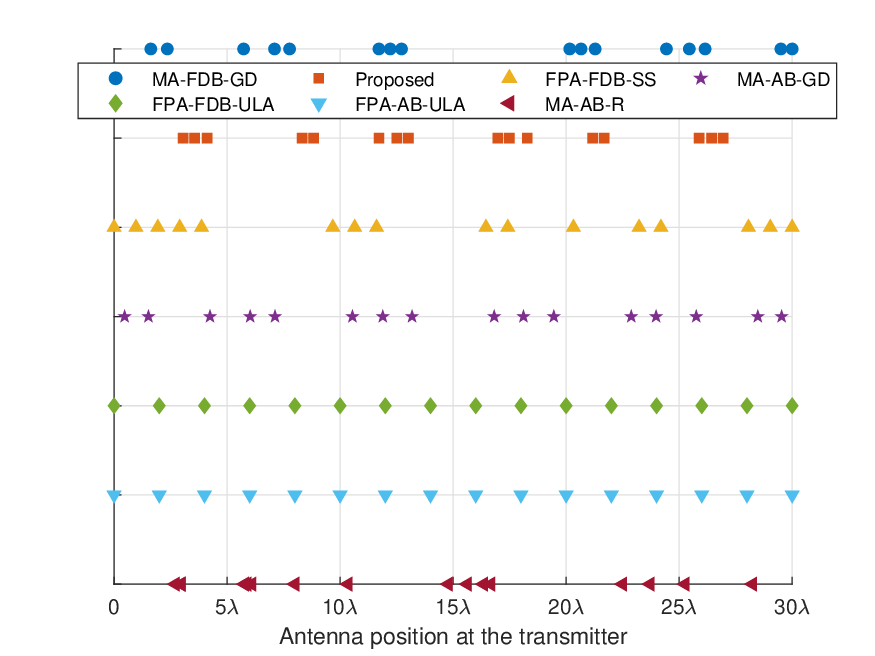}\\
  \caption{Comparison of antenna positions at the transmitter.}\label{ap}
  \end{center}
\end{figure}

\begin{figure}[t]
  \begin{center}
  \includegraphics[width=3in]{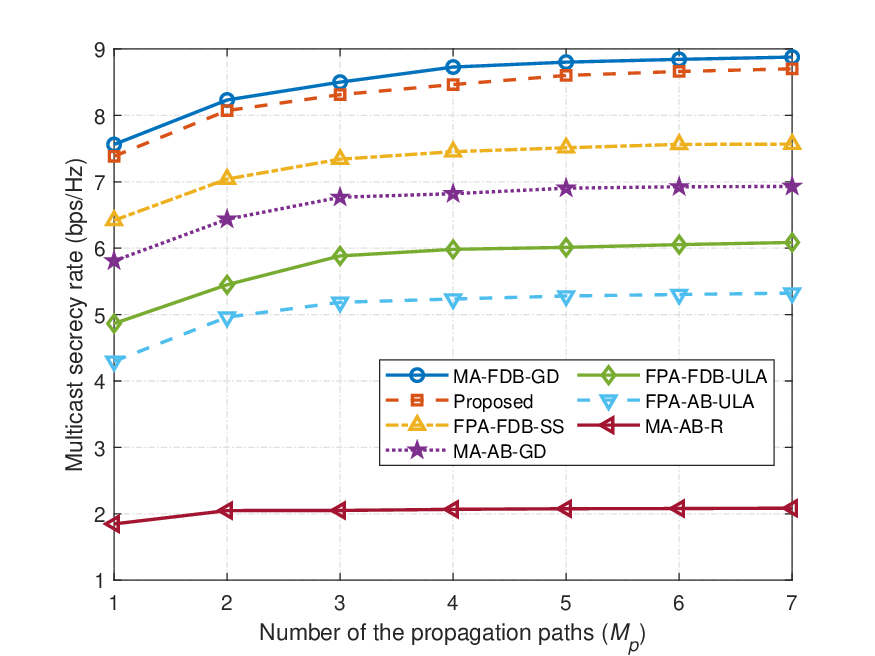}\\
  \caption{Comparison of MSR with varying numbers of propagation paths ($M_p$).}\label{ml}
  \end{center}
\end{figure}

Fig. \ref{ap} compares antenna positions across different architectures in a scenario with \( N_b = 4 \) and \( N_e = 4 \), where LUs are uniformly distributed within \([10^\circ, 30^\circ]\) and EVEs within \([70^\circ, 90^\circ]\). In MA-based methods, antennas are grouped into clusters, contrasting with the uniform distribution observed in FPA-based methods. These clusters allow the MA-enabled system to enhance channel information for LUs while suppressing it for EVEs, leading to significant MSR improvements. The results also demonstrate that random antenna placement (MA-AB-R) fails to achieve comparable MSR gains, highlighting the critical role of strategic antenna positioning in secure communication.

Fig. \ref{ml} shows the MSR of different schemes versus the number of propagation paths $(M_p)$ at the BS. It can be observed that the MSR of MA-based systems is significantly lower than that of FPA-based schemes due to the interference mitigation gain achieved through MA positioning optimization. Meanwhile, the proposed scheme with an AB architecture attains performance close to that of the MA-FDB-GD architecture employing FDB. In addition, the MSR of both the proposed and benchmark schemes increases with $M_p$. This is because, as the number of channel paths per user grows, spatial diversity improves and the correlation among channel vectors of different users decreases. Moreover, MA systems can exploit the pronounced channel variations to further reduce channel correlation, highlighting their potential to mitigate correlation effects and maintain high secrecy performance in dense networks.

\begin{figure}[t]
  \begin{center}
  \includegraphics[width=3in]{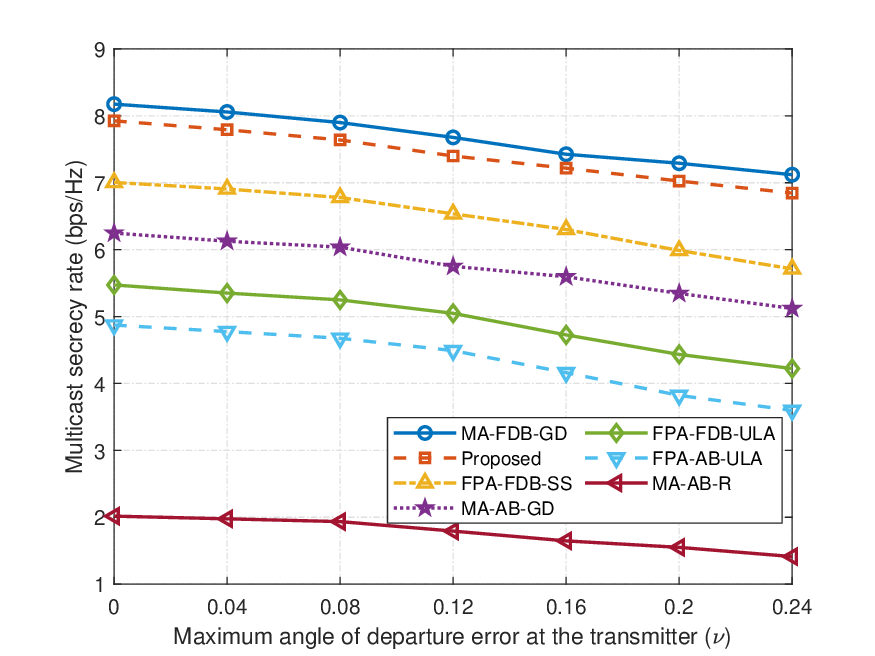}\\
  \caption{Comparison of MSR with varying values of the maximum angle of departure error at the transmitter ($\nu$).}\label{imcsi2}
  \end{center}
\end{figure}

\begin{figure}[t]
  \begin{center}
  \includegraphics[width=3in]{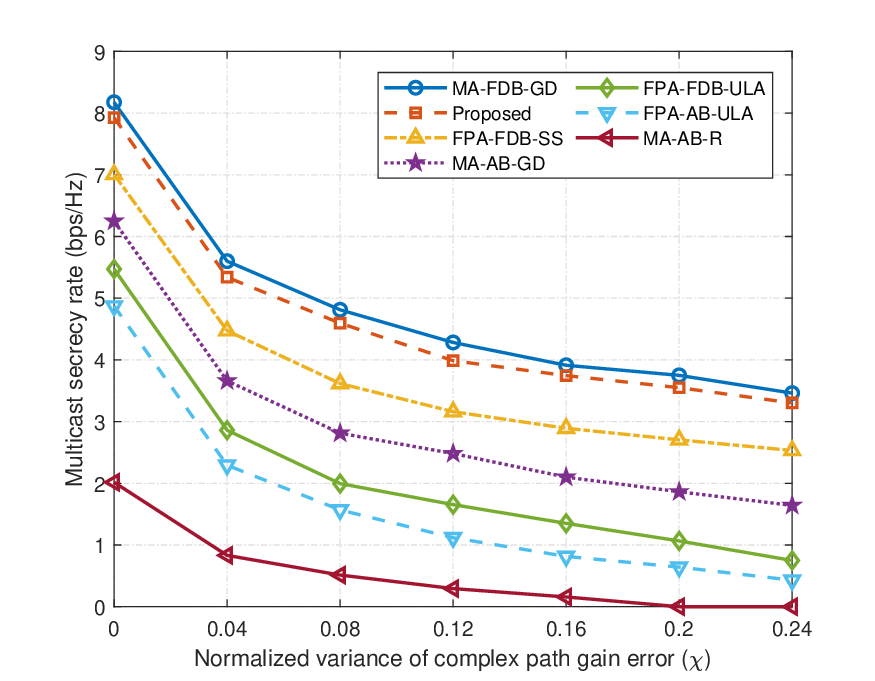}\\
  \caption{Comparison of MSR with varying values of the normalized variance of complex path gain error ($\chi$).}\label{imcsi1}
  \end{center}
\end{figure}

The results above assume perfect CSI at the BS. In practice, noise and limited training overhead make accurate CSI acquisition difficult, so it is necessary to assess how imperfect CSI affects the PLS performance of MA-enabled systems. Following \cite{zhu2023movable,ding2024movable}, we model CSI errors in the AoDs and complex path gains. Specifically, for the $m$-th path of link $i\in\{b,e\}$ (LU/EVE), let $\theta_{i,m}^{\mathrm{est}}$ and $\beta_{i,m}^{\mathrm{est}}$ denote the estimated AoD and complex path gain, respectively. The AoD error is modeled as, $\theta_{i,m}-\theta_{i,m}^{\mathrm{est}}\sim \mathcal{U}\!\left[-\tfrac{\nu}{2},\,\tfrac{\nu}{2}\right]$, where $\nu$ is the maximum AoD error. The normalized gain error is modeled as, $\frac{\beta_{i,m}-\beta_{i,m}^{\mathrm{est}}}{|\beta_{i,m}|}\sim \mathcal{CN}(0,\chi)$, where $\chi$ is the normalized error variance. As shown in Fig. \ref{imcsi2}, as the maximum AoD deviation $\nu$ increases, the MSR of all schemes decreases monotonically due to steering-vector mismatch, which reduces the beamforming gain toward LUs and increases unintended leakage to EVEs. Importantly, although this degradation trend is unavoidable, the proposed MA-AB scheme consistently maintains higher performance than the FPA-based baselines and the random-placement MA-AB (MA-AB-R), while remaining close to the MA-FDB-GD benchmark. This demonstrates that optimizing MA positions effectively preserves spatial degrees of freedom and LU/EVE separability under AoD uncertainty. Similarly, Fig. \ref{imcsi1} shows that as the normalized variance $\chi$ of the complex path gain error increases, the MSR of all schemes declines steadily, reflecting the combined amplitude and phase mismatch that lowers coherent array gain and aggravates leakage to EVEs. Nevertheless, throughout this degradation, the proposed MA-AB method retains a clear performance advantage over FPA-based baselines and closely tracks MA-FDB-GD, confirming that MA placement combined with analog phase control can decorrelate LU/EVE channels and stabilize secrecy performance even under gain uncertainty.

\section{Conclusion}
In this paper, we investigated the PLS of a MA-enabled MISO communication system with a hardware-efficient AB architecture. A multicast scenario was considered, where confidential information is transmitted to LUs while EVEs attempt to intercept the signal. To maximize the MSR, we jointly optimized the AB and MA positions under movement area and CM constraints. To address the NP-hard secrecy rate maximization problem, a PCPM framework was proposed. By transforming the inequality constraints into a penalty function and reformulating the problem as an unconstrained optimization on the PMS, we developed a PCGD algorithm for efficient updates of beamforming and antenna positions. Simulation results demonstrated that MA-enabled systems with AB achieve a well-balanced performance in terms of MSR and hardware cost. Future work will investigate two-timescale MA control, low-complexity partial-movement scheduling, and energy-/latency-aware actuator models, with the goal of reducing the energy overhead associated with real-time antenna movement.

\begin{appendices} 
\section{proof of lemma \ref{lemma1}}
\label{appendixA}
We demonstrate that problem (\ref{objectivefunction}) is NP-hard. To this end, we consider a special instance of problem (\ref{objectivefunction}).
First, we set the number of LUs and EVEs to $N_b = 1$ and $N_e = 1$, respectively. Furthermore, we fix the variable vector $\mathbf{p}$ to an arbitrary but specific constant vector, denoted as $\mathbf{\tilde{p}}$, which satisfies the original constraints on $\mathbf{p}$ (e.g., constraints (\ref{minidistance}) and (\ref{antennaselect})). This reduction strategy is valid because if a problem (such as (\ref{objectivefunction}) with variables $\mathbf{w}$ and $\mathbf{p}$) remains NP-hard when a subset of its variables (e.g., $\mathbf{p}$) is fixed, then the original problem is also NP-hard \cite{garey1979computers}.

Under these settings, ${\bf h}_b({\bf \tilde p})$ and ${\bf g}_e({\bf \tilde p})$ become constant vectors, denoted as ${\bf h}_0={\bf h}_b({\bf \tilde p})$ and ${\bf g}_0={\bf g}_e({\bf \tilde p})$, respectively. Problem (\ref{objectivefunction}) then simplifies to optimizing only over $\bf w$,
\begin{equation}
\begin{aligned}
\max _{{\bf w}} \quad &  \frac{ 1+{\hat t_b}|{\bf h}_0^H \mathbf{w}|^2}{  1+  {{\check t_e}|{\bf g}_0^H \mathbf{w}|^2}  } \\
\text { s.t. } \quad & \left|w_l\right|=1, \quad l=1, \ldots, L. \label{reduced_prob_p_fixed}
\end{aligned}
\end{equation}
Next, we consider a further special instance of problem (\ref{reduced_prob_p_fixed}) via choosing ${\bf g}_0=[1,0,\dots,0]$ and ${\check t_e}={\hat t_b}=1$. Then, we have ${\hat t_e}|{\bf g}_0^H \mathbf{w}|^2=|w_1|^2=1$ for any feasible $\bf w$, and thus problem (\ref{objectivefunction}) can be simplifies to,
\begin{equation}
\begin{aligned}
\max _{{\bf w}} \quad &  {\bf w}^H \left( {\bf h}_0{\bf h}_0^H  \right) {\bf w} \\
\text { s.t. } \quad & \left|w_l\right|=1, \quad l=1, \ldots, L. \label{lemma1e2}
\end{aligned}    
\end{equation}

We need the following lemma, which is established in the proof of Proposition 3.3 in \cite{zhang2006complex},
\begin{lemma} \label{lemma3}
Consider the following problem,
\begin{equation}
\begin{aligned}
\max _{{\bf w}} \quad &  {\bf w}^H {\bf Q} {\bf w}  \\
 \text { s.t. }\quad & \left|w_l\right|=1, \quad l=1, \ldots, L. \label{lemma1e3}
\end{aligned}    
\end{equation}
for some ${\bf Q}  \succeq 0$. Then, problem (\ref{lemma1e3}) is NP-hard in general.
\end{lemma}

Clearly, problem (\ref{lemma1e2}) is a special case of problem (\ref{lemma1e3}) with ${\bf Q} = {\bf h}_0 {\bf h}_0^H \succeq 0$. Since any instance of the NP-hard problem (\ref{lemma1e2}) is a special case of problem (\ref{reduced_prob_p_fixed}), we conclude that problem (\ref{reduced_prob_p_fixed}) is NP-hard. Consequently, the general problem (\ref{objectivefunction}) is also NP-hard. This completes the proof. $\hfill\blacksquare$

\section{Computation of the Euclidean gradients}
\label{appendixB}
To obtain the Euclidean gradients $\nabla_{{\bf w}} \phi({\bf w}, {\bf p})$ and $\nabla_{{\bf p}} \phi({\bf w}, {\bf p})$ with respect to ${\bf w}$ and ${\bf p}$, we first equivalently reformulate $\phi({\bf w}, {\bf p})$ in (\ref{reforoverpro}) as,
\begin{equation}
\begin{aligned}
\phi({\bf w}, {\bf p})= \frac{u_{e}({\bf w}, {\bf p})}{u_{b}({\bf w}, {\bf p})} + \varphi_p( {\bf p}),  \label{proofine} 
 \end{aligned}
\end{equation}
where,
\begin{subequations}
\begin{align}
&u_i({\bf w}, {\bf p}) = (-1)^{\epsilon_i} \alpha \log \sum\limits_{i=1}^{N_i} \varphi_i({\bf w}, {\bf p}), \quad i \in \{e, b\},\\ 
 &\varphi_p( {\bf p})= \rho\gamma  \left(\begin{array}{l} \textstyle\sum\limits_{l=1}^{L-1}  
 \log(1+e^{\frac{g_l(p_{l+1},p_{l})}{\gamma} }) \\+ \log(1+e^{\frac{f_1(p_{1})}{\gamma}})+ \log(1+e^{\frac{f_2(p_{L})}{\gamma}}) \end{array}\right),
\end{align}
\end{subequations}
and,
\begin{equation}
\varphi_i({\bf w}, {\bf p}) = \text{exp}\left((-1)^{\epsilon_i} \frac{1 + t_i|\sum_{m=1}^{M_p}{\beta_{i,m}}{\bf a}({\bf p},\theta_{i,m})\mathbf{w}|^2}{\alpha}\right), 
\end{equation}
where \( \theta_{i,m} \) represents \( \check\theta_{e,m} \) or \( \hat\theta_{b,m} \), $\beta_{i,m}$ represents \( \check\beta_{e,m} \) or \( \hat \beta_{b,m} \), \( t_i \) represents \( \check t_e \) or \( \hat t_b \), and \( i \) indexes \( e \) (EVE) or \( b \) (LU). The sign is determined by \( (-1)^{\epsilon_i} \), with \( \delta_i = 0 \) for \( i = e \) (positive) and \( \epsilon_i = 1 \) for \( i = b \) (negative). This notation will be used throughout the paper without further clarification. Next, we calculate the Euclidean gradients with respect to \( {\bf w} \) and \( {\bf p} \).

\textit{Calculation of} $\nabla_{{\bf w}} \phi({\bf w}, {\bf p})$: To obtain the Euclidean gradient related to ${\bf w}$, we reformulate (\ref{proofine}) by regarding ${\bf p}$ as a constant and ignoring the constant term, i.e.,
\begin{equation}
{\phi}({\bf w}, {\bf p}) = \frac{u_{e}({\bf w}, {\bf p})}{u_{b}({\bf w}, {\bf p})}.  
\end{equation}
According to the quotient rule, we have,
\begin{equation}
\begin{aligned}
\nabla_{{\bf w}} {\phi}({\bf w}, {\bf p}) &=\frac{\partial}{\partial {\bf w}}\left(\frac{u_{e}({\bf w}, {\bf p})}{u_{b}({\bf w}, {\bf p})}\right) \\&= \frac{\frac{\partial u_{e}({\bf w}, {\bf p})}{\partial {\bf w}}u_{b}({\bf w}, {\bf p}) -  \frac{\partial u_{b}({\bf w}, {\bf p})}{\partial {\bf w}}u_{e}({\bf w}, {\bf p})}{u^2_{b}({\bf w}, {\bf p})}.
\end{aligned}
\label{gw1}
\end{equation}
where,
\begin{equation}
\begin{aligned}
&\frac{\partial u_{i}({\bf w}, {\bf p})}{\partial {\bf w}} = \frac{\partial u_{i}({\bf w}, {\bf p})}{\varphi_i({\bf w}, {\bf p})}  \cdot \frac{\partial\varphi_i({\bf w}, {\bf p})}{\partial {\bf w}} = \frac{\alpha \cdot \frac{\partial\varphi_i({\bf w}, {\bf p})}{\partial {\bf w}}}{(-1)^{\epsilon_i} \sum\limits_{i=1}^{N_i} \varphi_i({\bf w}, {\bf p})} \\ 
&= \frac{\alpha \sum\limits_{i=1}^{N_i} \frac{\partial \left( (-1)^{\epsilon_i} \frac{t_i|{\sum_{m=1}^{M_p}{\beta_{i,m}}{\bf a}({\bf p},\theta_{i,m})}\mathbf{w}|^2 }{\alpha} \right)}{\partial {\bf w}} \cdot \varphi_i({\bf w}, {\bf p})}{(-1)^{\epsilon_i} \sum\limits_{i=1}^{N_i} \varphi_i({\bf w}, {\bf p})} \\ 
&= \frac{\sum\limits_{i=1}^{N_i} 2  t_i {(\sum_{m=1}^{M_p}|{\beta_{i,m}}|^2{\bf a}({\bf p},\theta_{i,m}) {\bf a}^H({\bf p},\theta_{i,m}))} \mathbf{w} \cdot \varphi_i({\bf w}, {\bf p})}{\sum\limits_{i=1}^{N_i} \varphi_i({\bf w}, {\bf p})},
\end{aligned}
\label{gw2}
\end{equation}
By combining (\ref{gw1})-(\ref{gw2}), the Euclidean gradient $\nabla_{{\bf w}} \phi({\bf w}, {\bf p})$ is obtained.

\textit{Calculation of} $\nabla_{{\bf p}} \phi({\bf w}, {\bf p})$: To compute the Euclidean gradient with respect to ${\bf p}$, we have,
\begin{equation}
{\nabla_{{\bf p}} {\phi}({\bf w}, {\bf p}) =\frac{\partial}{\partial {\bf p}}\left(\frac{u_{e}({\bf w}, {\bf p})}{u_{b}({\bf w}, {\bf p})}\right)+\frac{\partial}{\partial {\bf p}}\left(\varphi_p( {\bf p})\right).}\label{derp}
\end{equation}
For the term $\frac{\partial}{\partial {\bf p}}\left(\varphi_p({\bf p})\right)$, we have,
\begin{equation}
{\frac{\partial \varphi_p( {\bf p})}{\partial {\bf p}}= \rho  \left( \begin{array}{l} \textstyle\sum\limits_{l=1}^{L-1}  
\frac{g_l(p_{l+1},p_{l})}{1+e^{\frac{g_l(p_{l+1},p_{l})}{\gamma} }} \\+ \frac{f_1(p_{1})}{1+e^{\frac{f_1(p_{1})}{\gamma}}}+ \frac{f_2(p_{L})}{1+e^{\frac{f_2(p_{L})}{\gamma}}} \end{array}\right).} \label{penduse}
\end{equation}
We now turn to the derivation of the first term in (\ref{derp}), i.e., $\frac{\partial}{\partial {\bf p}}\left(\frac{u_{e}({\bf w}, {\bf p})}{u_{b}({\bf w}, {\bf p})}\right)$. We first define,
\begin{equation}
\begin{aligned}
&\Psi_i({\bf p}) = |\sum_{m=1}^{M_p}{\beta_{i,m}}{\bf a}({\bf p},\theta_{i,m})\mathbf{w}|^2,\\
&\psi_i({\bf p}) = \sum_{m=1}^{M_p}{\beta^*_{i,m}}{\bf a}^H({\bf p},\theta_{i,m})\mathbf{w}.
\end{aligned}
\end{equation}
Then, applying the quotient rule, we obtain,
\begin{equation}
\frac{\partial}{\partial {\bf p}}\left(\frac{u_{e}({\bf w}, {\bf p})}{u_{b}({\bf w}, {\bf p})}\right) = \frac{\frac{\partial u_{e}({\bf w}, {\bf p})}{\partial {\bf p}}u_{b}({\bf w}, {\bf p}) -  \frac{\partial u_{b}({\bf w}, {\bf p})}{\partial {\bf p}}u_{e}({\bf w}, {\bf p})}{u^2_{b}({\bf w}, {\bf p})},\label{dirp11}
\end{equation}
where $\frac{\partial u_i({\bf w}, {\bf p})}{\partial {\bf p}}$ is given by,
\begin{equation}
\begin{aligned}
\frac{\partial u_{i}({\bf w}, {\bf p})}{\partial {\bf p}}&{= \frac{\partial u_{i}({\bf w}, {\bf p})}{\varphi_i({\bf w}, {\bf p})}  \cdot \frac{\partial\varphi_i({\bf w}, {\bf p})}{\Psi_i({\bf w},{\bf p})}\cdot\frac{\partial\Psi_i({\bf w}, {\bf p})}{{\bf p}}}\\
&{= \frac{\sum\limits_{i=1}^{N_i} 2  t_i  \cdot \varphi_i({\bf w}, {\bf p})}{\sum\limits_{i=1}^{N_i} \varphi_i({\bf w}, {\bf p})}\cdot\frac{\partial\Psi_i({\bf w}, {\bf p})}{{\bf p}},}
\end{aligned}
\end{equation}
and the derivative of $\Psi_i({\bf w}, {\bf p})$ with respect to ${\bf p}$ is computed as,
\begin{equation}
\begin{aligned}
    &{\frac{\partial\Psi_i({\bf w}, {\bf p})}{{\bf p}}=\frac{\partial(\psi_i({\bf w}, {\bf p})\psi^*_i({\bf w}, {\bf p}))}{{\bf p}}}\\&{=\psi^*_i({\bf w}, {\bf p})\frac{\partial\psi_i({\bf w}, {\bf p})}{{\bf p}}+\psi_i({\bf w}, {\bf p})\left(\frac{\partial\psi_i({\bf w}, {\bf p})}{{\bf p}}\right)^*}\\&{=2\Re\left(\psi^*_i({\bf w}, {\bf p})\cdot\frac{\partial\psi_i({\bf w}, {\bf p})}{{\bf p}}\right),}
\end{aligned}    
\end{equation}
with,
\begin{equation}
{\frac{\partial\psi_i({\bf w}, {\bf p})}{{\bf p}} = 
\left[\frac{\partial\psi_i({\bf w}, {\bf p})}{\partial p_1}, \frac{\partial\psi_i({\bf w}, {\bf p})}{\partial p_2}, \dots, \frac{\partial\psi_i({\bf w}, {\bf p})}{\partial p_L}\right]^T.}
\end{equation}
Further, for any $l=1,...,L$, $\frac{\partial\psi_i({\bf w}, {\bf p})}{\partial p_l}$ can be derived as,
\begin{equation}
    \begin{aligned}
        {\frac{\partial\psi_i({\bf w}, {\bf p})}{\partial p_l}}&{=\sum_{m=1}^{M_p}{\beta^*_{i,m}}  e^{-j\tfrac{2\pi }{\lambda }\text{cos} \theta_{i,m}{ p_l}  }w_l}\\&{=-j\tfrac{2\pi }{\lambda }w_l\sum_{m=1}^{M_p}{\beta^*_{i,m}}\text{cos} \theta_{i,m}e^{-j\tfrac{2\pi }{\lambda }\text{cos} \theta_{i,m}{ p_l}  }}\label{dirp1end}
    \end{aligned}
\end{equation}
By combining equations (\ref{dirp11}) through (\ref{dirp1end}), we obtain the complete expression for $\frac{\partial}{\partial {\bf p}} \left(\frac{u_e({\bf w}, {\bf p})}{u_b({\bf w}, {\bf p})} \right)$. Together with the result from (\ref{penduse}), the Euclidean gradient $\nabla_{{\bf p}} \phi({\bf w}, {\bf p})$ is fully derived. This completes the proof. $\hfill\blacksquare$ 

\section{proof of theorem \ref{theorem1}}
\label{appendixC}
To establish the proof, we first demonstrate that the algorithm achieves a sufficient decrease in each iteration. Specifically, by applying the Armijo line search strategy in (\ref{Als}), we have,
\begin{equation}
\begin{aligned}
  {\phi({\bf w}_k, {\bf p}_k) - \phi({\bf w}_{k+1}, {\bf p}_{k+1})} & {\ge} \\ &{- \upsilon_k  \operatorname{grad}^H\phi({\bf w}_k, {\bf p}_k)  {\bf d}_{({\bf w}_k, {\bf p}_k)},} \label{therem2}
\end{aligned}
\end{equation}
where $\upsilon_k=\tau^r  {\hat \upsilon}$. Since the descent direction \({\bf d}_{({\bf w}_k, {\bf p}_k)}\) is constructed from the negative gradient components, there exists a constant \(\tilde{\tau} > 0\) such that,
\begin{equation}
  { - \operatorname{grad}^H\phi({\bf w}_k, {\bf p}_k)  {\bf d}_{({\bf w}_k, {\bf p}_k)} \geq {\tilde \tau} \| \operatorname{grad}\phi({\bf w}_k, {\bf p}_k) \|_2^2. }\label{therem3}
\end{equation}
Combining (\ref{therem2}) and (\ref{therem3}), we obtain,
\begin{equation}
  { \phi({\bf w}_k, {\bf p}_k) - \phi({\bf w}_{k+1}, {\bf p}_{k+1}) \geq c_{\text{dec}} \| \operatorname{grad}\phi({\bf w}_k, {\bf p}_k) \|_2^2,} \label{therem4}
\end{equation}
where $c_{\text{dec}} = \upsilon_k {\tilde \tau} >0$. According to (\ref{therem4}), we conclude that the algorithm achieves a sufficient decrease at each iteration.

We can now complete the proof. The proof is based on a standard telescoping sum argument. The desired inequality for all $k=0,1,\dots,K-1$ is obtained as follows,
\begin{subequations}
    \begin{align}
        \phi({\bf w}_0, {\bf p}_0) - &\phi_{\text{low}}  \ge \phi({\bf w}_0, {\bf p}_0) - \phi({\bf w}_K, {\bf p}_K) \\ & = \sum\limits_{k=0}^{K-1} \phi({\bf w}_k, {\bf p}_k) - \phi({\bf w}_{k+1}, {\bf p}_{k+1}) \\& \ge K c_{\text{dec}} \min _{k=0,1,\dots,K-1}||\operatorname{grad}\phi({\bf w}_k, {\bf p}_k) \|_2^2
    \end{align}
\end{subequations}
where $\phi_{\text{low}}=0$ is the lower bound value for the objective function. To get the limit statement, observe that $\phi({\bf w}_{k+1}, {\bf p}_{k+1}) \le \phi({\bf w}_k, {\bf p}_k)$ for all $k$ by (\ref{therem4}). Then, taking $K$ to infinity we see that,
\begin{equation}
    \phi({\bf w}_0, {\bf p}_0) -  \phi_{\text{low}} \ge \sum\limits_{k=0}^{\infty} \phi({\bf w}_k, {\bf p}_k) - \phi({\bf w}_{k+1}, {\bf p}_{k+1}),
\end{equation}
where the right-hand side is a series of nonnegative numbers. The bound implies that the summands converge to zero, thus,
\begin{equation}
\begin{aligned}
    0 &= \lim_{k \to \infty} \phi({\bf w}_k, {\bf p}_k) - \phi({\bf w}_{k+1}, {\bf p}_{k+1}) \\ &\le c_{\text{dec}} \lim_{k \to \infty}\|\operatorname{grad}\phi({\bf w}_k, {\bf p}_k) \|_2^2,
    \end{aligned}
\end{equation}
which confirms that $\|\operatorname{grad}\phi({\bf w}_k, {\bf p}_k) \|_2\to0$. Now, let $\{{\bf w},{\bf p}\}$ be a limit point of the sequence of iterates. By definition, there exists a subsequence of iterates
$\{{\bf w}_{(0)},{\bf p}_{(0)}\},\{{\bf w}_{(1)},{\bf p}_{(1)}\},\{{\bf w}_{(2)},{\bf p}_{(2)}\},\dots$ which converges to $\{{\bf w},{\bf p}\}$. Then, since the norm of the gradient of $ \phi({\bf w}, {\bf p})$ is a continuous function, it commutes with the limit and we find,
\begin{equation}
\begin{aligned}
    0 &= \lim_{k \to \infty}\|\operatorname{grad}\phi({\bf w}_k, {\bf p}_k) \|_2^2 = \lim_{k \to \infty}\|\operatorname{grad}\phi({\bf w}_{(k)}, {\bf p}_{(k)}) \|_2^2 \\ & = \|\operatorname{grad}\phi\left(\lim_{k \to \infty}( {\bf w}_{(k)}, {\bf p}_{(k)})\right) \|_2^2 = \|\operatorname{grad}\phi( {\bf w}, {\bf p}) \|_2^2,
\end{aligned}
\end{equation}
showing that all limit points generated by the sequence are stationary points. This completes the proof. $\hfill\blacksquare$

\end{appendices}

\bibliographystyle{IEEEtran}
\bibliography{Bibliography}

\vfill

\end{document}